\documentclass[useAMS,usenatbib]{mn2e}
\usepackage{amsmath}
\usepackage{graphicx}
\graphicspath{{./figs/}}
\def\beq{\begin{equation}}
\def\eeq{\end{equation}}
\def\barray{\begin{eqnarray}}
\def\earray{\end{eqnarray}}
\def\beqarray{\begin{eqnarray}}
\def\eeqarray{\end{eqnarray}}

\def\rmP{{\rm P}}

\newcommand{\Mpc}{\>{\rm Mpc}}

\newcommand{\mpch}{\>h^{-1}{\rm {Mpc}}}

\newcommand{\kB}{\>k_{\rm B}}

\newcommand{\msunh}{\>h^{-1}\rm M_\odot}

\def\gtsima{$\; \buildrel > \over \sim \;$}
\def\ltsima{$\; \buildrel < \over \sim \;$}
\def\prosima{$\; \buildrel \propto \over \sim \;$}
\def\gsim{\lower.7ex\hbox{\gtsima}}
\def\lsim{\lower.7ex\hbox{\ltsima}}
\def\simgt{\lower.7ex\hbox{\gtsima}}
\def\simlt{\lower.7ex\hbox{\ltsima}}
\def\simpr{\lower.7ex\hbox{\prosima}}
\def\la{\lsim}
\def\ga{\gsim}
\def\lta{\la}
\def\gta{\ga}

\newdimen\hssize
\newdimen\hdsize
\begin{document}
%

\title[Probing Hot Gas in Galaxy Groups]
      {Probing Hot Gas in Galaxy Groups through the Sunyaev-Zeldovich Effect}

\author[Ran Li et. al]
       {\parbox[t]{\textwidth}{
        Ran Li$^{1,2}$\thanks{E-mail:lir@bac.pku.edu.cn},
        H. J. Mo$^{2}$,
        Zuhui Fan$^{1}$,
        Frank C. van den Bosch$^{3,4}$,
        Xiaohu Yang$^{5}$}
        \vspace*{3pt} \\
  $^{1}$Department of Astronomy, Peking University, Beijing 100871, China\\
  $^{2}$Department of Astronomy, University of Massachusetts, Amherst
        MA 01003, USA\\
  $^{3}$Department of Physics and Astronomy, University of Utah, 115
        South 1400 East, Salt Lake City, UT 84112-0830, USA\\
  $^{4}$Astronomy Department, Yale University, P.O. Box 208101
        New Haven, CT 06520-8101, USA\\
  $^{5}$Shanghai Astronomical Observatory, the Partner Group of MPA,
        Nandan Road 80, Shanghai 200030, China}


\date{}

\pagerange{\pageref{firstpage}--\pageref{lastpage}}
\pubyear{2008}

\maketitle

\label{firstpage}


\begin{abstract}
  We investigate the potential of exploiting the Sunyaev-Zeldovich
  effect (SZE) to study the properties of hot gas in galaxy groups. It
  is shown that, with upcoming SZE surveys, one can stack SZE maps
  around galaxy groups of similar halo masses selected from large
  galaxy redshift surveys to study the hot gas in halos represented by
  galaxy groups.  We use various models for the hot halo gas to study
  how the expected SZE signals are affected by gas fraction, equation
  of state, halo concentration, and cosmology. Comparing the model
  predictions with the sensitivities expected from the SPT, ACT and Planck
  surveys shows that a SPT-like survey can provide stringent
  constraints on the hot gas properties for halos with masses $M\ga
  10^{13}\msunh$. We also explore the idea of using the cross
  correlation between hot gas and galaxies of different luminosity to
  probe the hot gas in dark matter halos without identifying galaxy
  groups to represent dark halos. Our results show that, with a galaxy
  survey as large as the Sloan Digital Sky Survey and with the help of
  the conditional luminosity function (CLF) model, one can obtain
  stringent constraints on the hot gas properties in halos with masses
  down to $10^{13}\msunh$. Thus, the upcoming SZE surveys should
  provide a very promising avenue to probe the hot gas in relatively
  low-mass halos where the majority of $L^*$-galaxies reside.
\end{abstract}


\begin{keywords}
galaxies: halos ---
galaxies: clusters: general ---
large-scale structure of Universe ---
dark matter ---
cosmology theory ---
methods: statistical
\end{keywords}


\section{introduction}

In the current cold dark matter (CDM) paradigm, the large-scale
structure in the Universe grows due to gravitational instability,
forming dark matter halos within which galaxies form. Since the cosmic
gas is initially cold and well mixed with the dark matter, the gas
component is expected to follow the evolution of the dark matter
component until it is heated by accretion shocks during halo
formation, resulting in extended gaseous halos that trace out the
potential wells of dark matter halos. In the simplest case where the
process is assumed to be non-radiative, the total amount of hot gas
contained in a gaseous halo is expected to be roughly the amount of
gas initially associated with the dark matter halo, and the
distribution of the gas is governed by hydrostatic equilibrium. In
reality, however, the situation is much more complicated. In addition
to radiative processes which cause the halo gas to cool and form
stars, effectively removing it from the hot phase, various feedback
processes, such as supernova explosions and AGN activity, may act as
an efficient source of reheating, strongly impacting the properties of
the hot gas. Consequently, the amount of hot halo gas and its
distribution may be very different from that expected from the simple
non-radiative model. Indeed, the total amount of hot gas in low-mass
halos, such as those hosting spiral galaxies, is observed to be much
lower than that expected from the universal baryon fraction. Even for
clusters of galaxies, where the hot gas fraction is observed to be
comparable to the expected universal fraction, the state and
distribution of the hot gas appear to be strongly affected by star
formation and AGN activity. Thus, the current state of hot gas in dark
matter halos reflects not only the outcome of gravitational collapse,
but also contains important information regarding galaxy formation and
evolution within dark matter halos. Therefore, the study of the hot
halo gas can provide important constraints on galaxy formation and
evolution in a way complementary to that provided by observation of
stars and cold gas.

The hot halo gas has traditionally been observed through its diffuse
X-ray emission. However, since the X-ray emissivity depends
sensitively on the gas density, this method is only powerful for hot
gas with relatively high density, such as that in the inner parts of
massive clusters/groups of galaxies. An alternative way to probe the
hot halo gas is through the Sunyaev-Zel'dovich effect (hereafter SZE)
the hot gas generates in the cosmic microwave background (CMB) due to
inverse-Compton scattering \citep{SZ72}.  Because CMB photons gain
energy as they traverse hot gas, the surface brightness of the CMB (in
a given band) in the direction of a cluster/group will be changed. In
the Rayleigh-Jeans tail (where $h_{\rm P}\nu/k_{\rm B}\ll T_{\rm
  CMB}\approx 2.73 {\rm K}$, with $\kB$ the Boltzmann constant and
$h_\rmP$ the Planck constant), the change is a decrement. This
observable decrement is proportional to the electron pressure
intergrated along the line of sight, and is more sensitive than X-ray
emission in probing hot gas within regions of relatively low density,
such as in the outskirts of dark matter halos.  Moreover, the SZE
``brightness'' (i.e. the change in CMB brightness due to the SZE) is
independent of redshift, making it possible to study the evolution of
hot gas in dark matter halos out to high redshift.

The first detection of the thermal SZE with high significance was
reported in 1978 \citep{Bir_et78}, six years after the concept was
proposed by \citet{SZ72}. Since then, technological advances have been
such that this method has become an important tool in probing the hot
gas in galaxy clusters \citep[e.g.][]{Myers_et97,Grego01,Nord09}. As
mentioned above, since the SZE surface brightness is independent of
redshift, it is also considered a powerful tool to detect and study
galaxy clusters at relatively high redshift, which can be used to
constrain the rate of structure formation in the Universe
\citep[e.g.][]{Birkinshaw99,Holder2001,Muchovej07}. In addition, the
SZE of galaxy clusters has been combined with X-ray observation
\citep[e.g.][]{Turner02,LaRoque06} and weak lensing
\citep[e.g.][]{bartelmann01,seal06,Umetsu09} to accurately probe
the shape and mass distribution of clusters of galaxies. 

Despite this progress, the application of the SZE is still in its
early stage, and the observational samples available today are still
small. However, this situation will change drastically in the near
future. New ground-based telescopes such as the South Pole Telescope
(SPT)\footnote{http://astro.uchicago.edu/spt/} and the Atacama
Cosmology Telescopy
(ACT)\footnote{http://www.physics.princeton.edu/act/index.html}, as
well as the Planck\footnote{http://astro.estec.esa.nl/Planck/}
satellite, are expected to detect thousands of clusters through the
SZE\citep[e.g.][]{Moodley08}. However, even with these new observations,
  only the hot gas profile of the
relatively massive systems (with masses larger than a few times $10^{14}\msunh$)
are expected to be studied individually. In the case of less massive
halos, which host the majority of bright galaxies in the Universe, the
data will be insufficient for individual detections. However, one may
bypass this restriction by stacking large numbers of groups together
in order to increase the signal-to-noise (hereafter S/N). Such an
analysis, however, requires a pre-selected sample of groups.
Fortunately, large surveys of galaxies such as the 2-degree Field
Galaxy Redshift Survey \citep[2dFGRS;][]{Colless01} and the Sloan
Digital Sky Survey \citep[SDSS;][]{York00} are now available for
selecting large and uniform samples of galaxy groups that cover a wide
range of halo masses
\citep[e.g.][]{Goto05,Miller05,Berlind06,Yang05,Y07,koester07}. This
makes it possible to study the average properties of hot halo gas in
relatively low-mass halos.

In this paper, we explore the potential of using stacks of galaxy
groups to probe and study the hot gas in relatively low-mass halos
through its SZE. We base our investigation on groups selected from
current galaxy redshift surveys combined with the expected SZE data
from on-going surveys such as Planck, ACT and SPT. We consider a
number of plausible models to describe the properties of the hot halo
gas, and compare our predictions with the detection limits expected
from ongoing surveys. In addition, we
also explore the possibility of probing the hot gas in low mass dark
matter halos by using the cross-correlation between SZE maps and
galaxies of different luminosities, which has the advantage that it
does not require a pre-identification of galaxy groups. 

The structure of this paper is as follows. In Section~\ref{sec:halo}
we present our models for dark matter halos and for the hot halo
gas. In Section~\ref{sec:SZE} we describe the SZE expected from dark
matter halos of given masses.  Section~\ref{sec:CLF} describes in
detail how to calculate both the cross-correlation between hot gas and
dark matter halos, and between hot gas and galaxies of a given
luminosity. In Section~\ref{sec:halopop}, we discuss the properties of
the group catalog to be used in our modeling.
Section~\ref{sec:result} contains our predictions of various models
and their detectabilities with the on-going SZE surveys such as SPT,
ACT, and Planck. We summarize our conclusions in Sec.\ref{sec:sum}

Unless specified otherwise, we adopt a $\Lambda$CDM cosmology with
parameters given by the WMAP 3-year data \citep[][hereafter `WMAP3
cosmology']{Spergel07}: $\Omega_{{\rm m},0}=0.238$,
$\Omega_{\Lambda,0}=0.762$, $h\equiv H_0/(100 \rm
{km\,s^{-1}\,Mpc^{-1}})=0.734$, and $\sigma_8=0.744$.

\section{Dark matter halos and hot gas}
\label{sec:halo}

\subsection{Dark matter halos}

We assume that dark matter halos follow the NFW \citep{NFW97} profile,
\begin{displaymath}\label{eq_NFW}
\rho_{\rm dm}(x) = \left \{ 
  \begin{array}{ll}
  \frac{\delta_0\rho_{\rm crit}}{x(1+x)^2} & \textrm{if \it x $\leq$ c} \\
                                       0 & \textrm{else}
  \end{array} 
\right . \,.
\end{displaymath}
Here $x \equiv r/r_{\rm s}$, with $r_{\rm s}$ a scale radius related
to the halo virial radius $r_{\rm vir}$ via the concentration
parameter, $c=r_{\rm vir}/r_{\rm s}$, and $\rho_{\rm crit}$ is the critical
density of the Universe. The characteristic over-density, $\delta_0$,
is related to the critical over-density of a virialized halo,
$\Delta_{\rm vir}$, by
\begin{equation}
\delta_0=\frac{\Delta_{\rm vir}}{3}\frac{c^3}{\ln(1+c) -c/(1+c)}\,.
\end{equation}
For the $\Lambda$CDM cosmology considered here, we adopt the form of
$\Delta_{\rm vir}$ given by \cite{Bryan98} based on the spherical
collapse model:
\begin{equation}
\Delta_{\rm vir}=18\pi^2+82[\Omega_{\rm m}(z)-1]-39[\Omega_{\rm m}(z)-1]^2\,,
\end{equation}
where $\Omega_{\rm m}(z)$ is the cosmological density parameter of
matter at redshift $z$. Thus, the concentration parameter $c$ is the
only parameter required to specify the density profile of a halo of
mass $M$. Numerical simulations show that halo concentration decreases
gradually with halo mass \citep[e.g.][]{Bul01,Eke01}. However the
exact mass-dependence of the concentration parameter has not yet been
well constrained, neither theoretically nor observationally. Various
fitting formulae based on numerical simulations have been proposed in
the literature \citep[e.g.][]{Bul01,Zhao03,Dolag04,Maccio07,Zhao09}.
For the purpose of this paper, the difference between these different
formulae does not affect our results qualitatively.  In what follows,
we adopt the fitting formula for $c=c(M,z)$ given by
\cite{Maccio07}. For simplicity we do not include the scatter in
$c$ for halos of a given mass at a given redshift.

\subsection{The Hot Gas Component}
\label{sec:hotgas}

By far the most common model used to describe the density
distribution, $\rho_g(r)$, of hot gas in clusters is the
$\beta$-model, which consists of a constant density core and falls off
as $\rho_g \propto r^{-\beta}$ at large radii. Recent observation,
however, has shown that the $\beta$-model is not a good approximation
for the gas density distribution outside the core regions of clusters
\citep[e.g.][]{Vikhlinin06,Kom01,Hallman07}. In this paper, we adopt a
model in which the gas is assumed to be in hydrostatic equilibrium(HE)
with a given equation of state. This assumption is supported by
numerical simulations \citep[e.g.][]{Evrard96,Bryan98,Thomas01}, and
has as an additional advantage that it is straightforward to link the
gas properties to those of the dark matter halos.

For an isothermal equation of state, the virial temperature of a halo
is defined as
\begin{equation}\label{Tvir}
T_{\rm vir} = {\mu m_{\rm p} \over 2 \kB}\,{G M \over r_{\rm vir}}\,,
\end{equation}
where $m_{\rm p}$ is the proton mass. We assume a fully ionized gas
with a mean molecular weight $\mu =0.588$. In reality, however, the
hot gas within dark matter halos is not expected to be isothermal.
Therefore, we consider two alternative models; one in which we assume
a more realistic, polytropic equation of state, and one in which we
consider some amount of initial entropy injection. In either case, the
density profile of the gas is obtained by solving the hydrostatic
equation,
\begin{equation}\label{eq:HE}
\frac{{\rm d}P_{\rm g}}{{\rm d}r} = 
-\frac{G\rho_{\rm g}(r)M(<r)}{r^2}\,,
\end{equation}
where $M(<r)$ is the total mass within radius $r$. For simplicity,
throughout this paper we will ignore the gravity of the gas.

\subsubsection{Polytropic Model}
\label{sec:polytrop}

Let us first consider the polytropic case, which we regard as our
fiducial model. The equation of state in this case is
\begin{equation}
P_{\rm g} \propto \rho_{\rm g}^{\Gamma}\,,
\end{equation}
with $\Gamma$ the polytropic index. Throughout this paper we adopt
$\Gamma = 1.2$, in agreement with both observations
\citep[e.g.][]{Fin_etal07} and with numerical simulations
\citep[e.g.][]{lewis00,Borgani04,Ascasibar03}.  With this equation of
state, it is straightforward to obtain the density profile $\rho_{\rm
  g}$ and the temperature profile $T_{\rm g}$:
\begin{equation}\label{poly_rho}
\rho_{\rm g}(x) = \rho_{{\rm g},0} \, \eta_{\rm poly}(x)\,,
\end{equation}
and
\begin{equation}\label{T_poly}
T_{\rm g}(x) = T_{{\rm g},0} \, \eta_{\rm poly}^{\Gamma-1}(x)\,,
\end{equation}
where, as before, $x=r/r_{\rm s}$, and $T_{{\rm g},0}$ and $\rho_{{\rm
    g},0}$ are the central temperature and central density,
respectively.  The function $\eta_{\rm poly}$ is dimensionless and is
obtained from the hydrostatic equation:
\begin{equation}
\eta_{\rm poly}(x) = \left[ 1 - B\,g(x)\right]^{1/(\Gamma-1)}\,,
\end{equation}
where
\begin{equation}
g(x) = 1-\frac{\ln(1+x)}{x},
\end{equation}
and
\begin{equation}
B = 2 \, \left(\frac{\Gamma-1}{\Gamma}\right)
  \left(\frac{T_{\rm vir}}{T_{{\rm g},0}}\right)
  \left(\frac{c}{f(c)}\right)\,,
\end{equation}
with
\begin{equation}
f(x) = \ln(1+x)-\frac{x}{1+x}\,.
\end{equation}
For given halo mass and redshift the density and temperature profiles
of the gas are therefore specified by three parameters: $\Gamma$,
$T_{{\rm g},0}$ and $\rho_{{\rm g},0}$. We set $\Gamma=1.2$, and the
other two free parameters are specified by boundary conditions.

As the first boundary condition, we assume that the temperature at the
virial radius is equal to the virial temperature, in good agreement with
numerical simulations \citep[e.g.][]{FWB99,Rasia04}. The central
temperature of the polytropic gas can then be written as
\begin{equation}\label{T_{g,0}}
T_{{\rm g},0} = T_{\rm vir}\,\left[ 1 + 2 \, {\Gamma-1 \over \Gamma} \,
{c \, g(c) \over f(c)}\right].
\end{equation}
To specify $\rho_{{\rm g},0}$, we fix the total gas inside the halo to be a
universal fraction of the halo mass. In particular, we assume that a
fraction $f_{\rm star}$ of the gas initially associated with the halo
either has formed stars or is in the cold phase.  The central density
of the hot gas is then given by
\begin{equation}\label{rho_g0}
\rho_{{\rm g},0} = (1-f_{\rm star}) \, {\Omega_{{\rm b},0} \over
\Omega_{{\rm m},0}} \left[ {4 \pi r_s^3 \over M} 
\int_0^c \eta_{\rm poly}(x) \, x^2 \, {\rm d}x \right]^{-1}\,,
\end{equation}
with $\Omega_{{\rm b},0}$ and $\Omega_{{\rm m},0}$ the cosmological
density parameters of baryons and of total matter, respectively.

asubsubsection{Entropy Injection Models}
\label{sec:entropyinj}

X-ray observations of galaxy groups and clusters indicate that there
may be an excess in gas entropy over that given by the polytropic
model in the inner regions of low temperature clusters and groups
\citep{PCN99,LPC00,Fin_etal07}.  This suggests that some
non-gravitational processes, such as supernova explosions and/or AGN
feedback, may have raised the entropy of the gas. In our second model,
we therefore consider an entropy injection model similar to that in
\citet{Moodley08,Vo_etal_02}. Throughout, we define the `specific entropy' of
the gas as
\begin{equation}\label{S_e}
S = T_{\rm g} \, n_{\rm g}^{-2/3}\,,
\end{equation}
where $n_{\rm g} = \rho_{\rm g}/(\mu m_{\rm p})$.  Ignoring the
details of the entropy injection process, we simply add a constant
entropy term, $S_{\rm inj}$, to the polytropic entropy profile,
$S_{\rm poly}$, which follows from Eq.~(\ref{S_e}) upon substitution
of Eqs.~(\ref{poly_rho}) and~(\ref{T_poly}). Defining $F_{\rm g}(r) =
M_{\rm g}(<r)/M_{\rm vir}$, with $M_{\rm g}$ the hot gas mass within
$r$, we model the final specific entropy distribution of the gas as
\begin{equation}\label{S_ent}
S_{\rm ent}(F_{\rm g}) = S_{\rm poly}(F_{\rm g}) + S_{\rm inj}\,.
\end{equation}
Based on X-ray observations of clusters \citep[e.g.][]{LPC00,PCN99},
we set the entropy floor to be either $100\,{\rm keV\,cm^2}$
(hereafter model `Ent100') or $200\,{\rm keV\,cm^2}$ (hereafter model
`Ent200'). The gas pressure is then $P_{\rm ent}(r)\propto S_{\rm
ent}\rho_{\rm ent}^{5/3}$, where $\rho_{\rm ent}$ is the gas density
profile.  In order to obtain $\rho_{\rm ent}(r)$, we solve the
following equations,
\begin{eqnarray} \label{eq:ode}
\frac{{\rm d}P_{\rm ent}}{{\rm d}r}&=&-\rho_{\rm ent}\frac{G M(<r)}{r^2},\\
\frac{{\rm d}F_{\rm g}}{{\rm d}r}&=&4\pi r^2\frac{\rho_{\rm ent}}{M_{\rm vir}},
\end{eqnarray}
by setting the boundary conditions as follows.  A natural boundary
condition is given by $F_{\rm g}(0)=0$. However, the choice of the other
boundary condition is less straightforward.  In the polytropic case
described above, we have set $F_{\rm g}(r_{\rm vir})= (1-f_{\rm star}) \,
(\Omega_{{\rm b},0}/\Omega_{{\rm m},0})$. However, this condition is
not expected to be valid in the presence of entropy injection, simply
because the increase in entropy changes the temperature in the inner
part of the halo, which may drive part of the gas out of the
halo. Here, somewhat arbitrarily, we choose to set the boundary
condition at $r_{\rm vir}$ by assuming that $T_{\rm ent}(r_{\rm vir}) =
T_{\rm vir}$. At least this model has the desirable property that it
reduces to the polytropic model described above in the limit $S_{\rm
inj}\to 0$. 
\begin{figure*}
  \begin{center}
    \includegraphics[width=\textwidth]{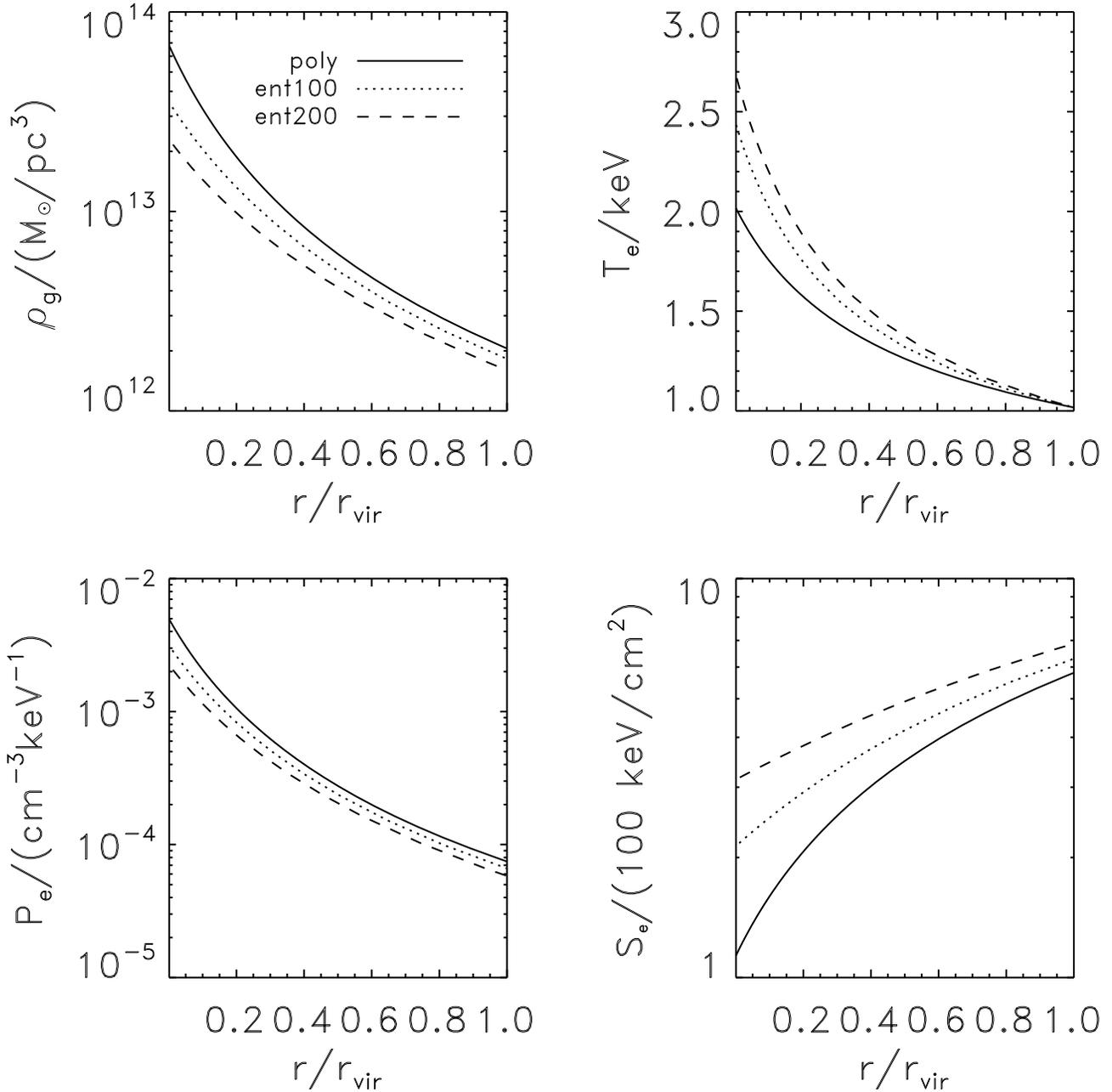}
  \end{center}  
\caption{Properties of the hot gas in a halo with mass
  $M=10^{14}\msunh$ at redshift $z=0.1$. Results are shown for all
  three gas models discussed in the text: the polytropic model, and
  the entropy injection models Ent100 and Ent200. In all three cases
  $f_{\rm star} = 0.1$. The various panels show the density (upper
  left-hand), temperature (upper right-hand), electron pressure (lower
  left-hand) and entropy (lower right-hand) of the gas, all as a
  function of halo-centric radius.}  \label{fig:gas}
\end{figure*}

\subsubsection{Physical Properties of the Hot Gas}
\label{sec:physprop}

In order to link the above model to the SZ effect generated by hot
electrons, we estimate the electron pressure as
\begin{equation}\label{Pe}
P_{\rm e} = \kB \, n_{\rm e} \, T_{\rm e}\,.
\end{equation}
The gas is assumed to be fully ionized, so that
\begin{equation}
n_{\rm e} = \frac{(1+\chi_{\rm H})}{2}\frac{\rho_{\rm g}}{m_{\rm p}} \,,
\end{equation}
with the hydrogen mass fraction $\chi_{\rm H}=0.76$. The electron
temperature is assumed to be the same as that of the gas, i.e.,
$T_{\rm e} = T_{\rm g}$.

\begin{figure}
\begin{center}
\includegraphics[width=0.5\textwidth]{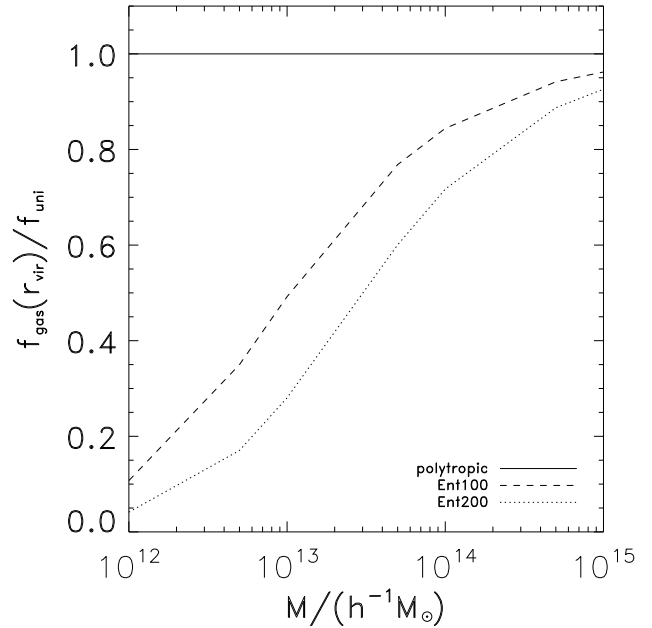}
\caption{The ratio between the gas fraction in a halo and the
  universal gas fraction as a function of halo mass. Results are shown
  for the polytropic model (solid line), for which this ratio is equal
  to unity by construction, and for the entropy injection models
  Ent100 (dashed line) and Ent200 (dotted line). Stronger entropy
  injection results in a reduction of the hot gas density at small
  radii.}
\label{fig:remain}
\end{center}
\end{figure}

Fig.\ref{fig:gas} shows the hot gas properties inside a halo of mass
$M = 10^{14}\msunh$ at redshift $z=0.1$.  The results are shown for
all three gas models discussed above: the polytropic model, Ent100 and
Ent200. In all cases the mass fraction in stars and cold gas is
assumed to be $0.1$ times the total baryon fraction (i.e. $f_{\rm
  star} = 0.1$).  Note that the polytropic model predicts a more
concentrated gas profile and a higher electron pressure profile than
the other two models.  As we will see, this results in a stronger
predicted total SZE signal. Entropy injection modifies the gas
profile, resulting in a higher temperature in the inner regions of the
halo.  To satisfy the HE within the same dark
matter halo, the gas density in the entropy injection models is
therefore reduced relative to the polytropic case, and part of the gas
now resides outside the (virial radius of the) halo. This is
illustrated in Fig.~\ref{fig:remain}, which shows the ratio between
the gas fraction in a halo and the universal gas fraction, as a
function of halo mass for the three models considered here. In the
case of a $10^{14}\msunh$ halo, this shows that entropy injection of
$100 {\rm keV cm^{-2}}$ ($200 {\rm keV cm^{-2}}$) reduces the hot gas
mass fraction inside the virial radius by $\sim 15\%$ ($\sim 30\%$),
compared to the polytropic case without entropy injection.

It is important to point out that the hot gas profile depends on the
boundary conditions adopted. Current simulations show that the gas
tends to trace the dark matter distribution in the outer
part \citep[e.g.][]{JS00,Yoshikawa,lewis00}, but the temperature
profile of the hot gas is still poorly understood. Unfortunately, the
boundary conditions for the hot gas are not well constrained by
observation, and different boundary conditions have been used in
literature.  For example, \cite{Kom01} derived a hot gas profile
(KS model, hereafter) assuming polytropic gas and that the 
density profile of the gas matches that of the dark matter 
in the outer regions of the halo, while \cite{OBB05} set the 
boundary condition by requiring the gas surface pressure to match an exterior
pressure. These different boundary conditions result in sizable
differences in gas profile, especially in the outer parts of the halo.
In our model, the hot gas distribution is assumed to have a sharp
cutoff at about the virial radius. However, in reality, the gas in the
vicinity of dark matter halos may also be heated by the collapse of
large scale structure, contributing to the SZE. Although we adopt the
boundary conditions described above, we will occasionally comment on
uncertainties that may arise from our specific choise of boundary
conditions.

\section{The expected SZE signals from dark matter halos}
\label{sec:SZE}

\subsection{The SZ effect}

The temperature fluctuations caused by the SZE can be written as
\begin{equation}\label{dt_T}
\frac{\Delta T}{T_{\rm CMB}} = \Theta_{\rm sz}(x) \, y_{\rm comp} - 
\beta\tau\,,
\end{equation}
where $T_{\rm CMB}\approx 2.73 \rm K$ \citep{Mat99}, and $x = h_{\rm
  P} \nu /\kB T_{\rm CMB}$, with $h_{\rm P}$ the Planck constant and
$\nu$ the photon frequency.  The first term on the right-hand side is
the thermal SZE caused by the thermal motions of electrons; the second
term is the kinematic SZE, caused by peculiar bulk motions along the
line of sight of the object in question. The thermal SZE has a
spectral shape given by
\begin{equation}\label{fsz}
\Theta_{\rm sz}(x) = x \frac{e^x+1}{e^x-1}-4\,,
\end{equation}
and a frequency-independent Compton parameter, $y_{\rm comp}$,
which is related to the projected electron pressure by:
\begin{equation}\label{yc}
y_{\rm comp} = \frac{\kB \sigma_{T}}{m_{\rm e}c^2}
\int n_{\rm e} T_{\rm e} {\rm d}l\,,
\end{equation}
with $\sigma_{T}$ the Thomson cross section. In the isothermal case,
\begin{equation}\label{yiso}
y_{\rm comp} = \frac{\kB T_{\rm e}}{m_{\rm e} c^2} \, \tau\,,
\end{equation}
with
\begin{equation}\label{tau}
\tau = \sigma_{T}\int n_{\rm e} {\rm d}l \,,
\end{equation}
the Thomson optical depth. The kinematic term is given by $\beta =
v_{\rm pec}/c$, where $v_{\rm pec}$ is the bulk peculiar velocity
along the line of sight. In what follows, we consider only the thermal
part of the SZE, treating the kinematic part as `contamination'.

\subsection{The SZ effect produced by halos of a given mass}

Before exploring the SZ effect produced by a sample of groups that
have similar masses, we first focus on the thermal SZE due to a single
dark matter halo.  Assuming spherical symmetry, the Compton parameter
$y_{\rm comp}$ around a halo of mass $M$ can be written in terms of
the `projected electron pressure', $\Sigma_{\rm p}$, as
\begin{equation}\label{yrm}
y_{\rm comp}(R|M) = \frac{\sigma_{T}k_{B}}{m_{\rm e}c^2}\Sigma_{\rm p}(R|M) \,.  
\end{equation}
We can express $\Sigma_{\rm p}$ in terms of the cross-correlation between
dark matter halos and the pressure profile of hot gas:
\begin{equation}\label{sig}
\Sigma_{\rm p}(R|M) = 2 \bar{n}_{\rm e}\int_{R}^{\infty} 
\xi_{\rm h,p}(r|M) {r \, {\rm d}r \over \sqrt{r^2 - R^2}} \,,
\end{equation}
with $R$ the projected distance to the halo center 

For convenience we will be working in Fourier space.  The cross
power spectrum between dark matter halos and the pressure field,
$P_{\rm h,p}(k|M)$, is related to the cross-correlation $\xi_{\rm 
h,p}(r|M)$ by
\begin{equation}
\xi_{\rm h,p}(r|M) = \frac{1}{2\pi^2} \int P_{\rm h,p}(k|M) 
\frac{\sin{kr}}{kr} k^2 {\rm d}k \,.
\end{equation}
This can be separated into two parts, a 1-halo term and a 2-halo term.
The 1-halo term describes the pressure profile of the halo's own hot
gas atmosphere. For a halo of a given mass $M$ at a given redshift, this
part of the power spectrum can be written as
\begin{equation}\label{p1h}
P_{\rm h,p}^{1h}(k|M) = \frac{1}{\bar{n}_{\rm e}} {\tilde u}_{\rm p}(k|M)\,,
\end{equation}
where ${\tilde u}_{\rm p}$ is the Fourier transform of the electron
pressure profile,
\begin{equation}\label{up}
{\tilde u}_{\rm p}(k|M) = \int_{0}^{r_{\rm vir}} 4\pi r^2 \frac{\sin{kr}}{kr} 
n_{\rm e}(r|M) T_{\rm e}(r|M) {\rm d}r\,.
\end{equation}
The 2-halo term describes the cross-correlation between halo centers
and the pressure distribution of hot gas in other halos. We assume
that on large scales the correlation function of halos is related to
that of the dark matter by a linear bias relation. The 2-halo term can
then be written as
\begin{equation}\label{p2h}
P_{\rm h,p}^{2h}(k|M) = \frac{P_{\rm lin}(k)}{\bar{n}_{\rm e}} b(M) 
\int_0^{\infty} {\tilde u}_{\rm p}(k|M') n(M') b(M') {\rm d}M'\,,
\end{equation}
where $P_{\rm lin}(k)$ is the linear power spectrum of the density
field, $n(M)$ is the halo mass function \citep[e.g.][]{PS74} and
$b(M)$ is the bias function for halos of mass $M$
\citep[e.g.][]{MW96,SMT01}.

The SZE brightness is the change in the CMB specific intensity 
caused by the SZE, and can be written as
\begin{equation}
\Delta I(x) = \zeta(x) I_{0} y_{\rm comp} \, ,
\end{equation}
where
\begin{equation}
I_{0}= \frac{2 h_{\rm P}}{c^2}
\left(\frac{\kB T_{\rm CMB}}{h_{\rm P}}\right)^3\,,
\end{equation}
\begin{equation}
\zeta(x) = \frac{x^4 e^x}{(e^x-1)^2} \left[x \frac{e^x+1}{e^x-1}-4\right]\,,
\end{equation}
and, as before, $x = h_{\rm P}\nu/\kB T_{\rm CMB}$. The integrated
Compton parameter, $Y$, can be written as:
\begin{equation}\label{Ytot}
Y = {1 \over D_A^{2}} \int y_{\rm comp}\, {\rm d}A\,,
\end{equation}
where ${\rm d}A = D_A^2 {\rm d}\Omega$, with $\Omega$ the solid angle
of the halo and $D_A$ its comoving angular-diameter distance.  In the
isothermal case, the value of $Y$ is determined by the gas temperature
and the total gas fraction, but is independent of the gas density
profile. The SZ flux density, the change in the CMB flux density 
caused by the SZE within the solid angle $\Omega$, can then be written 
in terms of the integrated Compton parameter $Y$ as
\begin{equation}
S_{\nu} = \frac{2 h_{\rm P}}{c^2}
\left(\frac{\kB T_{\rm CMB}}{h_{\rm P}}\right)^3 \, Y \, \zeta(x) \,,
\end{equation}
As one can see, the SZE brightness (flux) is proportional to the
Compton $y_{\rm comp}$ ($Y$) parameter. The SZ brightness is redshift
independent, but the total flux density of a cluster depends on its
redshift through $D_{A}$.

\section{Correlation between galaxies and hot halo gas}
\label{sec:CLF}

With a redshift survey of galaxies, another interesting quantity to
study is the cross correlation between galaxies of different
properties and hot halo gas. Since galaxy properties are tightly
correlated with the mass of their host dark matter halos, the
measurement of such correlation can be used to constrain the hot gas
distribution in dark matter halos as function of their mass. The correlation
between the galaxy density and thermal SZE has been studied in simulation and
early WMAP data\citep{Hern2004,Hern2006}. With upcoming SZE surveys, the
correlation between galaxies and hot gas can be studied more detailedly.
Here we
develop a formalism to study the SZE around galaxies in a given
luminosity bin.  The interpretation of such observation does not
require the identification of individual galaxy groups but is related
to the distribution of galaxy luminosities as a function of halo mass.
This connection between galaxy luminosity and halo mass is most
conveniently described by the conditional luminosity function
(hereafter CLF), $\Phi(L|M) {\rm d}L$, which specifies the average
number of galaxies with luminosities $L \pm {\rm d}L/2$ that reside in
a halo of mass $M$ \citep[][]{Yang03a,vandenbosch03}. For a given
cosmology, the CLF can be constrained using galaxy clustering
\citep[e.g.,][]{Yang03a,Cooray06,vdBosch07}, galaxy-galaxy lensing
\citep[e.g.,][]{Guzik02,LiRan09,Cacciato09}, satellite kinematics
\citep[e.g.,][]{vdBosch04,More09,More10} and galaxy group catalogues
\citep[e.g.,][]{Yang05b,Yang08}, or any combination thereof. In this
paper, we adopt the CLF parameterization motivated by the results
obtained from a large galaxy group catalogue \citep[see][]{Yang08}
with the parameters given in \citet{Cacciato09}, which have been
obtained using the combined constraints from the observed galaxy
luminosity function, the luminosity dependence of galaxy clustering,
and the SDSS group catalogue of \citet{Y07}.  As shown in
\citet{Cacciato09} and \citet{LiRan09} the same CLF model also
accurately matches the galaxy-galaxy lensing data of
\citet{Mandelbaum06}. Hence, in what follows we assume that the CLF is
accurately known, and we focus on how to use it in order to extract
information regarding the hot gas properties in dark matter halos
from the SZE-galaxy cross correlation.

The expected SZE signal around a given galaxy also depends on the
position of the galaxy within its host halo. For example, the SZE
around central galaxies, which reside at the centers of their host
halos, is expected to be quite different from that around satellite
galaxies located at off-center positions. Hence, it is convenient to
split the cross-power spectrum between galaxies and the hot gas
pressure into 4 parts. The 1-halo central term, the 1-halo satellite
term, the 2-halo central term, and the 2-halo satellite term:
\begin{eqnarray}\label{Pgp}
P_{\rm g,p}(k) & = & f_{\rm c} 
 \left[P_{\rm g,p}^{1h,\rm c}(k) + P_{\rm g,p}^{2h,\rm c}(k)\right] 
 \nonumber \\
              & + & f_{\rm s} 
 \left[P_{\rm g,p}^{1h,\rm s}(k) + P_{\rm g,p}^{2h,\rm s}(k)\right],
\end{eqnarray}
where $f_{\rm c}$ and $f_{\rm s}$ are the central and satellite
fractions, respectively, among among all galaxies in the luminosity
bin being considered. Note that the power spectrum depends on both the
redshift and luminosity of galaxies. For brevity, however, we will
not write down this redshift dependence explicitly.

In order to compute cross power spectrum of galaxies in a luminosity
bin $L\pm {\rm d}L/2$ we proceed as follows. Denote the probability
that a central (satellite) galaxy with luminosity $L$ resides in a
halo of mass $M$ by $\mathcal{P}_{\rm x}(M|L)$ [where ${\rm x}$
refers to either `c' (central) or `s' (satellite)]. Using Bayes'
theorem, we can write
\begin{equation}\label{PC}
\mathcal{P}_{\rm x}(M|L) = \frac{\Phi_{\rm x}(L\vert M) n(M)}
{\phi_{\rm x}(L)}\,,
\end{equation}
where $\Phi_{\rm x}(L|M)$ is the CLF for central (satellite) galaxies
in halos of mass $M$, and $\phi_{\rm x}(L) $ is the corresponding
luminosity function, which is related to the CLF according to
\begin{equation}\label{nbar}
\phi_{\rm x}(L) = \int_0^{\infty} \Phi_{\rm x}(L\vert M) 
n(M) \, {\rm d}M\,.
\end{equation}

Assuming that central galaxies always reside at the centers of their
host halos, we can write the 1-halo central term as
\begin{equation}\label{p1c_a}
P_{\rm g,p}^{1h,\rm c}(k\vert L) = \frac{1}{\bar{n}_{\rm e}}
\int_0^{\infty} \mathcal{P}_{\rm c}(M|L) {\tilde u}_{\rm p}(k|M)\,{\rm d}M \,.
\end{equation}
Inserting Eq.(\ref{PC}) into the above equation gives
\begin{equation}\label{P1c}
P_{\rm g,p}^{1h,\rm c}(k\vert L) =
  \frac{1}{\bar{n}_{\rm e} \phi_{\rm c}(L)} \int_0^{\infty}
  \Phi_{\rm c} (L\vert M) \, {\tilde u}_{\rm p}(k|M) \, n(M) \, {\rm d}M \,.
\end{equation}
To calculate the 1-halo satellite term, one needs to know how
satellites are distributed in their host halos. Here we make the
assumption that satellite galaxies follow a number density
distribution, $u_{\rm s}(r|M)$, that is similar to that of the dark
matter particles; i.e., $u_{\rm s}(r|M) \propto (r/r_{\rm
  s})^{-1}(1+r/r_{\rm s})^{-2}$.  The corresponding Fourier transform
is
\begin{equation}\label{us}
{\tilde u}_{\rm s}(k|M) = 4\pi\int_0^{r_{\rm vir}} u_{\rm s}(r|M)
  \frac{\sin(kr)}{kr}r^2 dr\,.
\end{equation}
The 1-halo satellite term can then be written as
\begin{equation} \label{P1s}
P_{\rm g,p}^{1h,\rm s}(k \vert L) = \frac{1}{\bar{n}_{\rm e}\phi_{\rm s} (L)}
  \int_0^{\infty} \Phi_{\rm s}(L\vert M) {\tilde u}_{\rm s}(k|M) 
  {\tilde u}_{\rm p}(k|M) \, n(M) \,{\rm d}M \,.
\end{equation}

The 2-halo term describes the correlation between galaxies and the hot
gas pressure in halos other than their own host halos, and can be
written as
\begin{equation}\label{P2h}
P_{\rm g,p}^{2h,\alpha}(k\vert L) = P_{\rm lin}(k)
    {\cal I}_{\rm x}(L) {\cal I}_{M} \,,
\end{equation}
where $P_{\rm lin}(k)$ is the linear power-spectrum of the density field,
and
\begin{equation}\label{Ic}
{\cal I}_{\rm c}(L) = \int_0^{\infty} \frac{\Phi_{\rm c}(L\vert M)}{\phi_{\rm c}(L)} 
\, b(M) \, n(M) \, {\rm d}M \,,
\end{equation}
\begin{equation}\label{Is}
{\cal I}_{\rm s}(L) = \int_0^{\infty} \frac{\Phi_{\rm s}(L\vert M)}{\phi_{\rm s}(L)}
 {\tilde u}_{\rm s}(k|M) \, b(M) \, n(M) \, {\rm d}M \,,
\end{equation}
and
\begin{equation}\label{Im}
 {\cal I}_M=\frac{1}{\bar{n}_e}\int^{\infty}_0 u_p(k|M) b(M) n(M) dM \,.
\end{equation}

In practice, we consider the signal produced by galaxies in a finite
luminosity bin, $[L_1, L_2]$, which can be obtained by integrating
the $L$-dependent quantities over $L$, and replacing $\Phi_{\rm x}(L|M)$ 
in the above equations by
\begin{equation}
\langle N \rangle_{\rm x}(M) = \int_{L_1}^{L_2} \Phi_{\rm x}(L|M) {\rm d}L\,.
\end{equation}

\section{Data Issues}
\label{sec:halopop}

In this section we discuss various characteristics of the on-going
SZE surveys that are needed in order to make realistic predictions
of the signal to be expected from the two different analyses proposed
here: stacking galaxy groups and cross-correlating galaxies with the
SZ signal.

\subsection{Group Catalogues and SZ Surveys}
\label{sec:cats}

One of the analyses we are proposing is to stack large numbers of
galaxy groups in order to probe the hot gas in relatively low mass
haloes. Galaxy group catalogues are best obtained from large galaxy
redshift surveys, and in what follows we will focus on the SDSS. The
wide sky coverage of the SDSS allows one to identify thousands of
galaxy groups within a redshift of about $0.15$ (see below). The
typical angular size of a $10^{14}\msunh$ halo is about 10 arcmin at
$z=0.1$. Hence, many of the SDSS groups are expected to have
sufficiently large angular sizes to be resolved in future SZE surveys
with typical resolutions of sub-arcmin to a few arcminutes.

Various groups have constructed galaxy group catalogues using the SDSS
\citep[e.g.][]{Goto05,Miller05,Berlind06,koester07,Yang05,Y07}. The
most suitable one for our purpose is the one published recently by
\citet[][hereafter Y07]{Y07}. This group catalog (hereafter GCY07) was
obtained from the SDSS Data Release 4 \citep[DR4;][]{Adelman06} using
the halo-based adaptive group finder developed by \citet{Yang05}.
Each group is assigned a halo mass according to the total stellar
mass it contains. As demonstrated in Y07, with the help of large mock
redshift surveys, this group finder works well not only for rich
groups but also for poor systems, facilitating studies of galaxy
groups that cover a wide range in halo masses.

In what follows, the GCY07 is used to demonstrate the feasibility of
the analysis we are proposing here.  In particular, we use the
GCY07 to estimate the number of groups in each mass bin and the
corresponding observational noise level.
\begin{figure}
  \begin{center}
    \includegraphics[bb=0 312 600 647,width=0.5\textwidth]{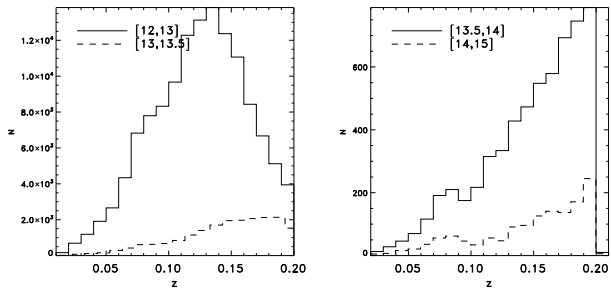}
  \end{center}
  \caption{The redshift distribution for group of different halo
    masses in the group catalog. In the left panel, the solid line
    shows the redshift distribution of halos in the range
    $[10^{12},10^{13}]$ $\msunh$, and the dashed line shows that for
    halos with masses in $[10^{13},10^{13.5}]$ $\msunh$. In the right
    panel, the solid line and dashed line represent halos of
    $[10^{13.5},10^{14}]$ $\msunh$, and $[10^{14},10^{15}]\msunh$,
    respectively.}
 \label{fig:reddis}
\end{figure}

Fig.\ref{fig:reddis} shows the redshift distribution of groups in
different mass ranges.  It is encouraging that the GCY07 provides a
fairly large number of groups in the low-redshift Universe, which
allows us to analyze groups in relatively narrow mass bins. For
instance, there are 114 groups with masses between $10^{13.9}$ and
$10^{14.1}$ $ \msunh$ in the redshift range $0.08$ - $0.11$. Table
\ref{tab:halo} lists the number of groups in different halo-mass and
redshift bins.  In the following sections, we will use the number of
groups in these samples as our basis for predicting the detectability
of the SZE in upcoming surveys.
\begin{table*}
\caption{The number of halos (groups) in different mass bins and redshift 
  bins in the group catalogue of Y07, which is based on the  SDSS DR4. The 
  first row lists the mass bins in $\log[M/(\msunh)]$. The first column 
  indicates the various redshift bins. }
\begin{tabular}{|c|c|c|c|c|c|}
 \hline
 & $[12.9,13.1]$ & $[13.4,13.6]$ & $[13.9,14.1]$ & 
   $[14.4,14.6]$ & $[14.9,15.1]$ \\
\hline
$[0.05 , 0.08]$ &  762 &  252 &  72 & 18 & 0 \\
$[0.08 , 0.11]$ & 1368 &  448 & 114 & 16 & 0 \\
$[0.11 , 0,14]$ & 2836 &  824 & 169 & 16 & 0 \\
$[0.14 , 0.17]$ & 3707 & 1208 & 270 & 38 & 2 \\
$[0.17 , 0.2] $ & 3864 & 1594 & 431 & 77 & 7 \\
\hline
\end{tabular}
\label{tab:halo}
\end{table*}

Table~\ref{table:t1} shows some important characteristics of the SPT,
ACT and Planck surveys.  For Planck, the coverage will be all-sky so
that the number of groups available for stacking is limited by the
optical survey. In this case, the numbers listed in
Table~\ref{tab:halo} can be directly used for computing the predicted
SZ signal. For the SPT survey, however, the aimed coverage is about $\sim
4000$ square degrees.  More importantly, it has zero overlap with the
SDSS DR4. However, the 2dFGRS, which has a depth similar to the SDSS,
has an overlap of about 1000 square degrees with the SPT.  On-going
optical surveys, such as the Dark Energy Survey (DES) \footnote
{https://www.darkenergysurvey.org/}, and the next generation galaxy surveys, such as
PanSTARRS\footnote{http://pan-starrs.ifa.hawaii.edu/} and
LSST\footnote{http://www.lsst.org/} will provide much larger and
deeper optical samples covering all, or at least part of, the SPT
survey area. In principle, the number of groups for stacking is then
limited by the SPT survey area which is similar to that of the SDSS
DR4. We would then expect the number of groups available in the same
mass and redshift ranges to be similar to those listed in Table
\ref{tab:halo}.  However, these surveys will only yield photometric
redshifts, which are far less reliable than the spectroscopic
redshifts available in, for example, the SDSS. Hence, group finders
specifically designed for redshift surveys, such as the halo-based
group finder of \citet{Yang05} used here, are not expected to perform
accurately on these surveys. However, in recent years some group
finders have been developed specificly for photometric surveys, and it
has been demonstrated that they can achieve high completeness and
purity \citep[e.g.][]{koester07, milkeraitis10}. Although these
methods are mainly restricted to massive groups and clusters, this is
not necessarily an important restriction for SZE studies such as those
proposed here, which, as we demonstrate below, are only able to probe
hot gas in relatively massive halos. In this work, we therefore make
the optimistic assumption that future surveys will ultimately provide
data in the SPT area of sufficient quality, so that our modeling based
on the GCY07 is relevant.
\begin{table}
  \caption{Characteristics of Planck, SPT and ACT surveys. 
    Data adapted from \citet{Bartlett06},\citet{Ho2009} and
    \citet{swetz09}}
\begin{tabular}{|c|c|c|c|c|}
\hline
Name & Freq.  & Res. FWHM & Inst. noise   & Survey Area \\
     &[GHz]   & [arcmin]  & [$\mu$K/beam] & [deg$^2$]\\
\hline
\hline
Planck & 143 & 7.1 &   6 & 40000 \\
       & 217 &   5 &  13 &       \\
       & 353 &   5 &  40 &       \\
\hline
SPT    & 150 &   1 &  10 &  4000 \\
       & 220 & 0.7 &  60 &       \\
       & 275 & 0.6 & 100 &       \\
\hline
ACT    & 148 & 1.4 &  15 &   2000 \\ 
       & 218 & 1.3 &     &        \\
       & 227 & 0.9 &     &        \\
\hline
\end{tabular}
\label{table:t1}
\end{table}

For comparison, we also explore the detectability with the ACT
survey. This survey has a beam size and instrumental noise similar to
the SPT, but a slightly smaller survey area. If the corresponding optical
survey has a similar depth, the expected number of groups in the ACT
area will be about 50\% of that in the SPT area. Thus, the signal to
noise ratio expected from the ACT is about 70\% of that from SPT.

\subsection{Noise and Contamination}
\label{sec:noise}

In order to examine the level of SZE that can be observed with the
samples described above, we need to take account of noise and signal
contamination in the observations. For the problem we are considering
here, the main source of noise will be instrument noise, while there
are three types of astrophysical sources that may cause signal
contamination: (i) the primary CMB anisotropy, (ii) point sources, and
(iii) the SZE produced by unresolved background clusters. We now
discuss each of these in turn.

The instrument noise (in $\mu$K/beam) expected for the various SZ
surveys is listed in Table~\ref{table:t1}. Throughout we will assume
that the instrument noise for different galaxy groups is uncorrelated,
so that stacking lowers the noise by a factor of $\sqrt{N}$, with $N$
the number of groups in the stack.

The contamination by the primary CMB anisotropy is expected to 
dominate at large scales. The noise level due to the primary  
is about 100 $\mu K$ per beam, which is much larger than the 
instrument noise. Fortunately, the primary anisotropy is frequency 
independent, while thermal SZE varies with frequency and vanishes 
at around 217 GHz. Thus, the primary contamination can in principle 
be separated from the SZE by using multi-band observations. 
\citet{Plagge2010} studied the SZE profile of galaxy clusters 
in the SPT survey. They used the 220 GHz map to subtract the 
background contamination and produced a set of band-subtracted 
map with a depth less than 20 $\mu K$. This provides hope
that the contamination due to the primary CMB can be properly 
subtracted. In this paper, this contamination is ignored.

Bright point sources such as quasars and star forming galaxies can
contaminate the SZE map on small scales in the form of bright spots.
However, since such sources are expected to be masked out, we do not
consider them. Un-resolved point sources, on the other hand, can
produce a significant contamination. In the case of unresolved IR
point-sources the contamination level is expected to be comparable to
the instrument noise \citep{WhiteMajumdar04}. Since the correlation
among these sources is not expected to play a significant role
\citep[e.g.][]{WhiteMajumdar04}, we can treat this contamination as
un-correlated noise.

Another class of unresolved point-sources that may be an important
source of contamination are radio galaxies. Many investigations
\citep[e.g.][]{WhiteMajumdar04, Staniszewski09, Plagge2010} have 
shown that the contamination by unresolved radio sources is lower than
that of the IR sources and unlikely to be an important contribution 
to the SZE noise. However, these sources may be correlated with 
the clusters and groups under investigation. In principle, 
such correlation can be understood with observation in wave-bands 
that are not sensitive to SZE. Recently
\citet[][]{Hall2010} and \citet[][]{Staniszewski09} argued 
that the clustering amplitude of radio galaxies is only a few 
percent of the mean background on arcminute scales. Thus, 
the contribution from clustered radio sources is at
a level of a few tenth $\mu K$. Such noise is not important 
for groups with masses $\sim 10^{14} \msunh$ or larger. 
However, since this noise does not decrease with stacking 
more groups, it may significantly affect results for 
groups of $\sim 10^{13} \msunh$ and needs to be included. 
  
In this paper, we treat the point source contamination as un-correlated
noise and account for it by simply doubling the instrument noise.

Previous investigations have demonstrated that the contamination by
the SZE background can be significant. For example, using light cones
constructed from an adiabatic hydrodynamical simulation,
\citet{Hallman07} found that unresolved halos and filaments can
contribute about $30\%$ to the total flux of a SZE map, and that there
is a significant chance (60\%) for a single beam to contain multiple
sources in a survey with a beam size as large as that of the Planck
survey. More recently, \citet{shaw08} studied this effect using a
similar method.  They derived a fitting formula for the SZE background
fluctuations and estimated their impact on the $Y$-$M$ relation.  They
found that the SZE background contamination is about 10 to 30 percent
of the total SZE flux of a cluster at low redshift. In order to
estimate the uncertainties introduced by this contamination we compute
the SZE angular power spectrum using our fiducial model (see Appendix),
from which we construct a mock all-sky map of SZE temperature
fluctuations, using the HEALPIX
package\footnote{http://healpix.jpl.nasa.gov} \citep{Gorski05}. The
angular resolution of our mock SZE map is $0.73\,{\rm arcmin}^{2}$ per
pixel (i.e., we set the HEALPIX parameter $N_{\rm side} = 4096$).
Assuming that the mean contribution of the background sources can be
subtracted, we can then estimate the {\rm rms} of the fluctuations in
the background on a given angular scale.  On 1-arcmin scale, we find
an $rms$ of $1.03 \times 10^{-6}\,{\rm arcmin}^2$ for the $Y$
parameter, in agreement with the results of \citet{shaw08}.  Unlike
the instrument noise, the noise generated by this contamination is
expected to be correlated on small scale. To take such correlation
into account, we mimic observations separately for the SPT, ACT and
Planck surveys. For Planck, we choose random directions in the mock
sky and estimate the noise within solid angles chosen to match the
virial radii of halos of different masses.  For SPT and ACT, we
estimate the noise in annuli around different random directions, with
the sizes of the annuli chosen to match the radial bin sizes to be
used.  The SZE background fluctuation around clusters at two different
directions is assumed to be uncorrelated. Hence, when stacking the
signal from multiple directions (i,e., multiple groups), the noise
contribution due to this contamination source decreases with
$\sqrt{N}$, where $N$ is the stacking number. This noise is added into
our error budget by assuming that it is independent of the other noise
sources.

To summarize, our noise model consists of instrument noise, which we
have artificially doubled to mimic the contribution of unresolved
IR point sources and Radio point sources(assumed to be uncorrelated),
   plus the contamination due to SZE background fluctuations obtained
   from our mock all-sky maps as described above.

\section {Results}
\label{sec:result}

\subsection{Fiducial Model}
\begin{figure}
\includegraphics[width=0.5\textwidth]{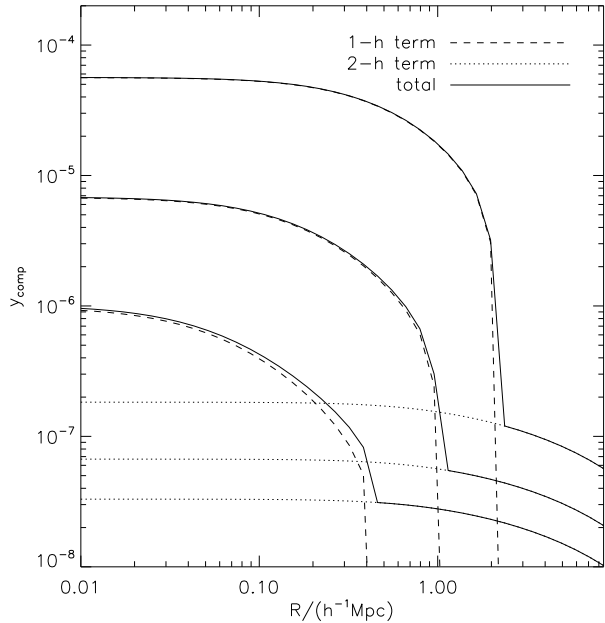}
\caption{The Compton parameter $y_{\rm comp}$ as a function of the
  projected halo-centric distance. The contributions to the $y$
  parameter are divided into 1-halo term and 2-halo term.  The three
  sets of lines from high to low represent the results for halos at
  redshift $z=0.1$ and with masses $10^{15}\msunh$, $10^{14}\msunh$,
  and $10^{13}\msunh$, respectively.}
  \label{fig:yfid}
\end{figure}
\begin{figure}
\includegraphics[width=0.5\textwidth]{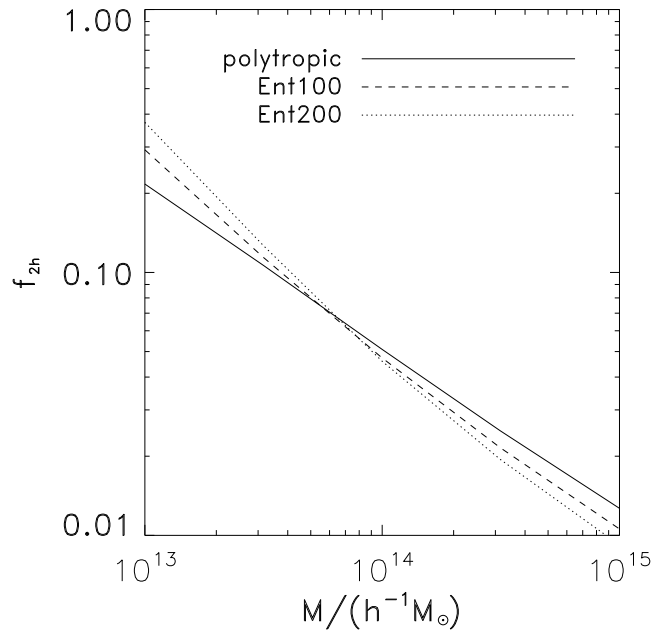}
\caption{The contribution of the 2-halo term to the total integrated
  Compton parameter, $Y$, within a projected radius equal to the halo
  virial radius as a function of halo mass for halos at $z=0.1$. The
  solid, dashed and dotted lines are results for the fiducial model,
  Ent100 and Ent200, respectively.}
    \label{fig:2h}
\end{figure}

We first show the SZE for our fiducial model, where the gas is assumed
to be in HE in NFW dark matter halos and to have
a polytropic equation of state with $\Gamma=1.2$.  The cosmological
parameters adopted are those from the WMAP3 data \citep{Spergel07}.
Fig.\ref{fig:yfid} shows the Compton parameter $y_{\rm comp}$ as a
function of the projected halo-centric distance $R$. Results are shown
for halos with three masses: $10^{13}$, $10^{14}$, and
$10^{15}\msunh$, respectively. The 1-halo and 2-halo terms are plotted
separately using different line styles. While both terms decrease
monotonically, the 1-halo term clearly dominates the SZE within the
virial radius. The 2-halo terms for all the three halo masses have the
same shape, and the difference in the amplitude is due to the mass
dependence of the linear halo bias, $b(M)$.  For halos with $M =
10^{13} h^{-1}M_{\odot}$, the two halo term becomes noticeable at the
outer part of the halo, because the 1-halo term is relatively
low. Note, however, that we have assumed that the hot halo gas only
extends out to the virial radius. If the hot halo gas extends out to
larger radii, the 1-halo term may still be important at larger radius.
 In particular, the models considered here ignore possible
contribution from the warm-hot intergalactic medium (WHIM) associated
with the filamentary and sheet-like structures in which the dark
matter halos are embedded. As a simple test of the potential impact of
such a WHIM component, we modified our fiducial model such that the
hot gas profile extends to two times the virial radius. This boosts
the SZE on scales around the virial radius by a factor $\sim 2$,
making it easily detectable for groups more massive than
$10^{13}\msunh$. This suggests that the analysis proposed here is also
very promising in probing the relatively elusive WHIM in the direct
vicinity of dark matter halos

In Fig.\ref{fig:2h} we plot the fractional contribution of the 2-halo
term to the total integrated Compton parameter $Y$ within the
projected radius $R=r_{\rm vir}$ as a function of halo mass. Results
are shown for all three gas equations of state considered here, as
indicated.  While the 2-halo contribution is negligible for the
most massive groups (i.e., $\lta 5\%$ for groups with $M \gta 10^{14}
h^{-1}M_{\odot}$), its contribution can reach as much as $\sim 40\%$
in haloes of $\sim 10^{13} h^{-1}M_{\odot}$.  As shown in
Fig.\ref{fig:gas}, entropy injection decreases the gas pressure inside
dark matter halos, causing the amplitude of the SZE to decrease for
both the 1-halo and 2-halo terms.  Since the impact of entropy injection
is more pronounced in less massive halos, and since the 2-halo term
reflects the contribution from halos of all masses, entropy injection
has a weaker effect on the 2-halo term than on the 1-halo term for low
mass halos, while the opposite applies to massive halos.

\subsection{Dependence on Model Parameters}
\label{sec:model}

\begin{figure*}
\begin{center}
  \includegraphics[width=1\textwidth]{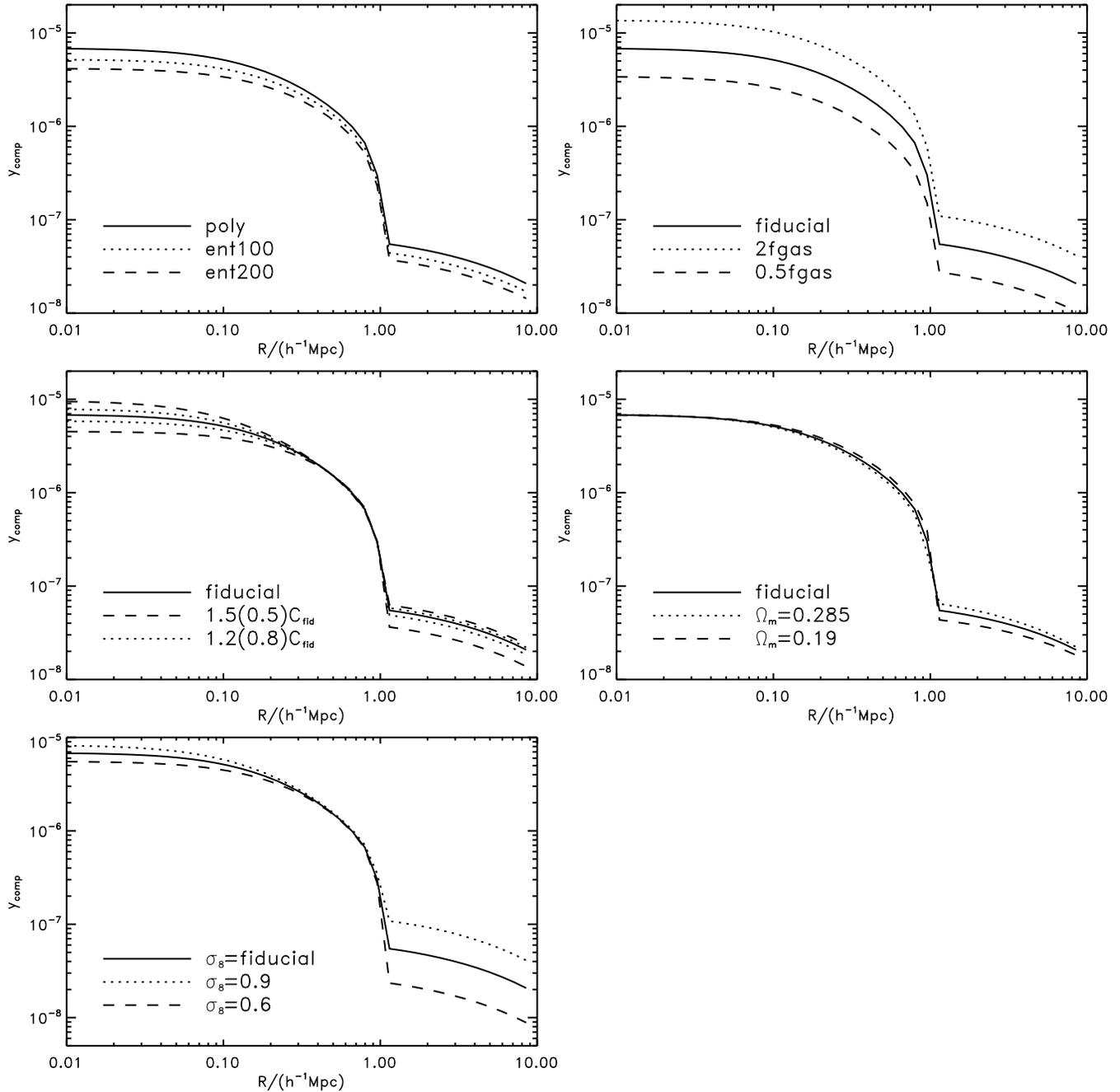}
\end{center}
\caption{The model dependence of the Compton parameter $y_{\rm
    comp}$. Here the halo has a mass of $10^{14}\msunh$ and is located
  at redshift $z=0.1$. The top left panel shows the effect of changing
  the gas equation of state. The top right shows the effect of
  changing gas fraction. The middle left panel shows the effects of
  changing the halo concentration by 20\% and 50\% . The middle right
  and bottom left panels show the impact of changing the matter
  density parameter, $\Omega_m$, and the power-spectrum normalization,
  $\sigma_8$, respectively, by 20\% relative to the values for our
  fiducial (WMAP3) cosmology.}
  \label{fig:ymod}
\end{figure*}

Since many of the parameters in our model are still poorly
constrained, we now investigate the effects on the predicted SZE
signal from changing the following model parameters: the equation of
state of the hot halo gas, the mass fraction of the hot gas, the
concentration of dark matter halos, and cosmological parameters. For
brevity, we only present results for a halo with $M = 10^{14} h^{-1}
M_{\odot}$ at redshift $z = 0.1$. Fig.\ref{fig:ymod} shows the Compton
parameter $y_{\rm comp}$ as a function of the projected halo-centric
distance $R$. The top left panel shows the dependence on the gas
equation of state. Here we compare the polytropic model with the two
entropy injection models, Ent100 and Ent200 (see Section
\ref{sec:hotgas} for details). Compared to the polytropic case, the
gas pressure profiles of the entropy models are lower. This is easy to
understand, as energy injection heats the gas at the halo center,
reducing the density of the gas in the central regions. As expected,
the reduction is more pronounced for a larger value of $S_{\rm
  inj}$. Note that the 1-halo term in the entropy models extends to a
slightly larger radius than in the polytropic model, which is due to
the fact that the energy injection drives part of the hot gas out of
the virial radius. The difference between the polytropic model and
Ent200 is within a factor of two, and the effect is more important for
lower-mass halos where the gravitational potential well is
shallower. Note that the impact of an entropy floor is also evident at
large radii, where the 2-halo term dominates, which is due to the
additional heating reducing the hot gas pressure in individual halos.

The top right panel of Fig.\ref{fig:ymod} shows the model dependence
on the hot gas fraction of the halo. As expected, $y_{\rm comp}$ is
directly proportional to the gas fraction, making it trivial to scale
our results for other choices of the gas fraction.  

Another model ingredient that remains somewhat uncertain are the halo
concentrations. Observationally, these are only poorly constrained
\citep[e.g.][]{comerford07,oguri09}, which is why one typically
resorts to the results from numerical simulations. These reveal a
scatter in concentration parameter, $c$, of about 30\% at a given halo
mass \citep[e.g.][]{Jing00}. More importantly, the mean halo
concentration is found to decrease with increasing halo mass
\citep[e.g.][]{Bul01,Zhao03,Dolag04,Maccio07,Zhao09}, but the exact
slope and normalization of this mass-dependence remain fairly
uncertain, with different authors claiming relations that are
significantly different. The middle left panel shows the effect of
changing the concentrations of the dark matter halos by $20\%$ (dotted
lines) and $50\%$ (dashed lines) relative to the fiducial
value. Clearly, halo concentrations can have a significant impact on
the SZE at small radii (and in the 2-halo term). Halos with larger
(smaller) concentrations are more (less) centrally concentrated, which
causes an increase (decrease) in the density of the hot gas at small
radii.

The middle right and bottom left panels of Fig.~\ref{fig:ymod} show
the effects of changing the cosmological parameters $\Omega_{\rm m}$
and $\sigma_8$, respectively.  Recall that for our fiducial model we
adopt the WMAP3 cosmology (see Table\ref{tab:cos}).  In $\Lambda$CDM
cosmologies, the 2-halo term, which reflects the clustering amplitude
of halos, depends on both $\Omega_{\rm m}$ and $\sigma_8$ through the
halo mass function and halo bias function, while the 1-halo term
depends on these parameters through the halo
concentrations. Fig.~\ref{fig:ymod} shows the impact on $y_{\rm
  comp}(R)$ of $20\%$ changes in the values of $\Omega_{\rm m}$ and
$\sigma_8$ relative to their fiducial values. As one can see, the SZE
increases with $\sigma_8$, while the dependence on $\Omega_{\rm m}$ is
rather weak. On scales dominated by the 1-halo term (i.e., $R \lta
r_{\rm vir}$), the SZE depends only weakly on $\Omega_{\rm m}$ and
$\sigma_8$. However, on larger scales, where the 2-halo term
dominates, the effects are larger. In particular, an increase
(decrease) of $\sigma_8$ by 20\% results in an increase (decrease) of
$y_{\rm comp}$ on large scales by a factor of $\sim 2$. A similar
change in $\Omega_{\rm m}$ only affects $y_{\rm comp}$ at the 13\%
level.
\begin{table}
\caption{Cosmological parameters used in the paper.}
\begin{tabular}{|c|c|c|c|c|c|c|}
 \hline & $\Omega_{{\rm m},0}$ & $\Omega_{\Lambda,0}$ & 
          $\Omega_{{\rm b},0}$ & $h$ & $n$ & $\sigma_8$ \\
  \hline
  \hline
WMAP1 & 0.3  & 0.7  & 0.040   & 0.7   & 1.0   & 0.9 \\
WMAP3 &0.238 &0.762 & 0.041   & 0.734 & 0.951 & 0.744\\
 \hline
\end{tabular}
\label{tab:cos}
\end{table}

\subsection{Predictions for the Planck Survey}

\begin{figure*}
  \begin{center}
    \includegraphics[width=\textwidth]{planck.eps}
  \end{center}
  \caption{The integrated Compton parameter $Y$ as a function of halo
    mass. The halo redshifts are set to $z=0.1$.  The left panel shows
    the results assuming different gas equations of state, while the
    right panel shows the results assuming different gas fraction. The
    green errorbars represent the 3 $\sigma$ instrument noise of the
    Planck telescope at 143 GHz, and the red errorbars shows the 3
    $\sigma$ error expected from the stack of groups in the
    corresponding mass range.}
  \label{fig:Yplk}
\end{figure*}
With the spatial resolution of Planck, low-mass halos are not
spatially resolved, and even the highest mass halos will only be
marginally resolved \citep{Aghanim97}. In this case, observation can
only be used to estimate the integrated Compton parameter $Y$ of an
entire group (or stack thereof). Fig.\ref{fig:Yplk} shows $Y$ as a
function of group mass, $M$. Here $Y$ is obtained by integrating
$y_{\rm comp}$ within the virial radius of the halo. As discussed in
Section \ref{sec:SZE}, for a group with a given mass, $M$, the
$Y$-parameter decreases with the redshift of the group.  For each mass
bin we therefore average $Y$ using the mass and redshift distributions
of the groups in GCY07.

To predict the error in a given mass bin, we use the number of groups
in the GCY07 (see table~\ref{tab:halo}) under the assumption that the
noise for each group is independent of that of other groups. In that
case the noise of the stacked signal decreases as $N^{-1/2}$, where
$N$ is the number of groups in the stack.

In the left-hand panel of Fig.\ref{fig:Yplk}, the $Y$ parameter is
shown for our 3 different models for the gas equation of state.  In
all three cases, the $Y$ parameter is almost the same at the high-mass
end.  For low mass groups, however, the entropy-injection models
predict lower values for $Y$. For example, for halos with $M \sim
10^{13}\msunh$ the $Y$ value predicted by the polytropic model is
larger than that of the Ent200 model by a factor of about two. In the
right-hand panel of Fig.\ref{fig:Yplk}, we show the effect of gas
fraction. Here again the results are shown for gas fraction that is 2
and $0.5$ times that adopted in the fiducial model. The larger error
bars show the 3 $\sigma$ detection limit for individual halos expected
for Planck.  The smaller error bars represent the uncertainties in the
corresponding stacks of groups.  Clearly, by stacking groups of
similar masses, the total SZE flux around groups with masses $M \gta
10^{13.5}\msunh$ can be constrained well. In particular, the GCY07 is
large enough to constrain the hot gas fractions to an accuracy of
$\lta 40\%$ for groups with masses higher than $10^{13.5}\msunh$.
However the sensitivity and spatial resolution of Planck is not
sufficient to distinguish among the different gas equations of state
considered here, at least not when using a group catalogue the size of
the GCY07. Larger and deeper optical surveys will be required to
reduce the noise to sufficiently low levels. For example, if the
number of groups is increased by a factor of 4, the observation will
be able to distinguish Ent200 and the fiducial model at $\sim 3\sigma$
confidence level.

\subsection{Predictions for the SPT and ACT Surveys}
\label{sec:spt}

\begin{figure*}
\begin{center}
 \includegraphics[width=\textwidth]{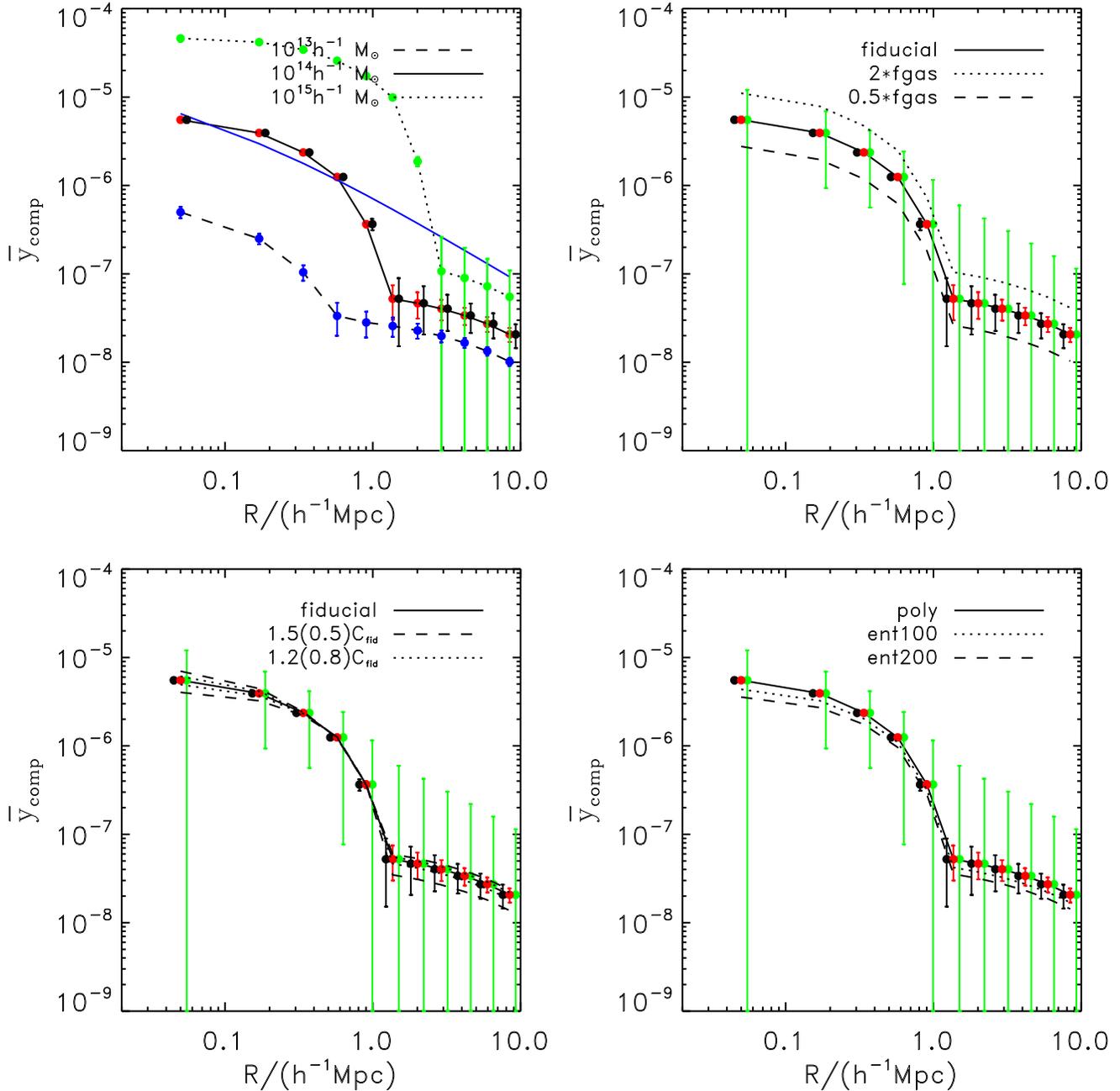}
 \caption{The average Compton parameter $\bar{y}_{\rm comp}$ as a
   function of the projected halo-centric distance $R$.  The top left
   panel shows the results of the fiducial model for different halo
   mass.  The errorbars show the expected $1\sigma$ uncertainty of the
   SPT for the corresponding stacks.The expected 1 $\sigma$
   uncertainties of observing one single group is shown using blue
   solid line.  The top right panel shows the results assuming
   different gas fraction. The bottom right panel shows the results
   assuming different gas equations of state, and the bottom left
   panel shows the result for different halo concentrations. In these
   three panels, we adopt a halo mass of $10^{14}\msunh$. The results
   are compared with SPT $1\sigma$ detection limit (green error bars)
   for a single observation, and the detection limit for stacks (red
   error bars). We also plot the $1\sigma$ detection limit of
     ACT(black error bars) for comparison. }
 \label{fig:Yspt}
\end{center}
\end{figure*}

In the case of the SPT, the resolution at 150 GHz is 1 arcmin with an
instrument sensitivity of 10 $\mu$K/beam. For comparison, the angular
sizes corresponding to the virial radii of groups with masses of
$10^{13}$, $10^{14}$ and $10^{15}$ $h^{-1}M_{\odot}$ at $z=0.1$ are
$5.24$, $11.3$ and $24.3$ arcmin, respectively. Thus, SPT has
sufficient spatial resolution to probe the actual hot gas profiles of
SDSS groups.  However, in order to obtain sufficient S/N, one needs to
stack groups of similar masses together, especially at the low mass
end. As for Planck, we base our analysis on galaxy groups in the
GCY07, even though the area on the sky covered by this group catalogue
(obtained from the SDSS DR4) has no overlap with the area covered by
the SPT survey (see discussion in Section~\ref{sec:cats}). For a group
of a given mass at a given redshift, we calculate the SZE profile and
smooth it over a one-arcmin scale using a top-hat window.  We then
average the signal in each mass bin according to the host halo mass
distribution given by the SDSS group catalog.

The estimate is made in 11 logarithmic bins of halo-centric distance
($R$), from the halo center to a maximum of $10 h^{-1}{\rm Mpc}$. To
maximize the S/N, we stack groups in a wide redshift range, from
$0.05$ to $0.2$, following the redshift distribution of the GCY07
group catalog. For each $R$-bin, we compute the average signal as
\begin{equation}\label{eq:spt}
\bar{y}_{\rm comp} = \frac{\Sigma_i w_i y_{i,{\rm comp}}}{\Sigma_i w_i} \,,
\end{equation}
where $y_{i,{\rm comp}}$ is the SZE signal of the $i$th group, and
$w_i$ is a weighting function chosen to be
\begin{equation}\label{weight}
w_i = \frac{1}{D_i^2} \,,
\end{equation}
with $D_i$ the angular diameter distance of the $i$th group.  This
weighting scheme is chosen to account for the fact that for haloes of
the same mass, the SZE flux within an annulus of a certain angular
distance is inversely proportional to the square of the distance of
the halo.  The noise of $\bar{y}_{\rm comp}$ in a bin of $R$ is
estimated through
\begin{equation}
\sigma_{\rm y} (R) = 
\sqrt{\frac{\Sigma_i w_i \, \sigma_{{\rm y},i}^2(R)}{\Sigma_i w_i}} \,,
\end{equation}
where $\sigma_{{\rm y},i}(R)$ is the measurement noise of the $i$th
group.  Assuming that the noise contributions from different beams
(i.e., different groups) are independent, we can estimate
$\sigma_{{\rm y},i}(R)$ through
\begin{equation}
\sigma_{{\rm y},i}(R) =\frac{\sigma_{\rm m}}{\sqrt{N_{i}(R)}}\,,
\end{equation}
where $\sigma_{\rm m}$ is the measurement noise of $\bar{y}_{\rm
  comp}$ per beam, and $N_{i}(R)$ is the number of beams the annulus
contains. Note that $N_{i}(R)$ depends on $i$ because the beam size
corresponds to different real space area at different redshifts. The
SZE background fluctuation, $\sigma_{\rm bg}$, is estimated as
described in Section~\ref{sec:noise}, and is added in quadrature to
$\sigma_{\rm y}$ to get the total uncertainty
\begin{equation}
\sigma_{\rm tot}=\sqrt{\sigma_{\rm bg}^2(R)+\sigma_{\rm y}^2(R)} \,.
  \label{eq:sigtot}
\end{equation}

The upper left-hand panel of Fig.\ref{fig:Yspt} shows the average
Compton parameter, $\bar{y}_{\rm comp}$, as a function of the projected
halo-centric distance $R$ for our fiducial model. Results are shown for
the three mass bins around $10^{13}$, $10^{14}$, and $10^{15}\msunh$,
respectively (see Table~\ref{tab:halo}).  The expected $1\sigma$
uncertainties for the stacks of groups are shown with the
errorbars. For comparison, the 1-$\sigma$ uncertainty expected from a
single group is indicated by the solid blue line.  This shows that the
SPT can only map the SZE profile of {\it individual} halos with masses
significantly larger than $10^{14}\msunh$. However, by stacking the
signal from groups in a catalogue the size of GCY07, one can study the
(average) hot gas distribution around halos with masses as low as
$\sim 10^{13} h^{-1}{\rm M}_{\odot}$. The remaining three panels of
Fig.\ref{fig:Yspt} show the predicted, average SZ profile for the
stack of GCY07 groups in the $10^{14} \msunh$ mass bin for different
gas mass fractions (upper right-hand panel), different gas equations
of state (lower right-hand panel), and different halo concentrations
(lower left-hand panel).  Comparing these model predictions with the
expected SPT sensitivity, it is clear that SPT can provide stringent
constraints on the properties of hot gas in dark matter halos.  Even
the 2-halo term can be detected with a group catalogue the size
of GCY07, allowing one to probe hot gas in the infall regions around
group-sized haloes (i.e., possibly associated with the WHIM inside
filaments and pancakes).

We also predict the signal and detection limit expected from ACT. The
procedure of the calculation is the same as for SPT, except that the
predicted signal is now smoothed with a 1.4 arcmin top-hat window to
match the resolution of ACT at 148 GHz. Since the signals are binned
into relative large annuli, the results expected from ACT and SPT are
almost identical, except that the signal-to-noise is slightly lower for ACT.
For comparison, we plot the error-bars expected from the ACT in
Fig.\ref{fig:Yspt}. As one can see, ACT is still able to provide
constraints on the hot gas fractions in halos with $M \gta
10^{14}\msunh$.
\begin{figure*}
  \begin{center}
    \includegraphics[width=\textwidth]{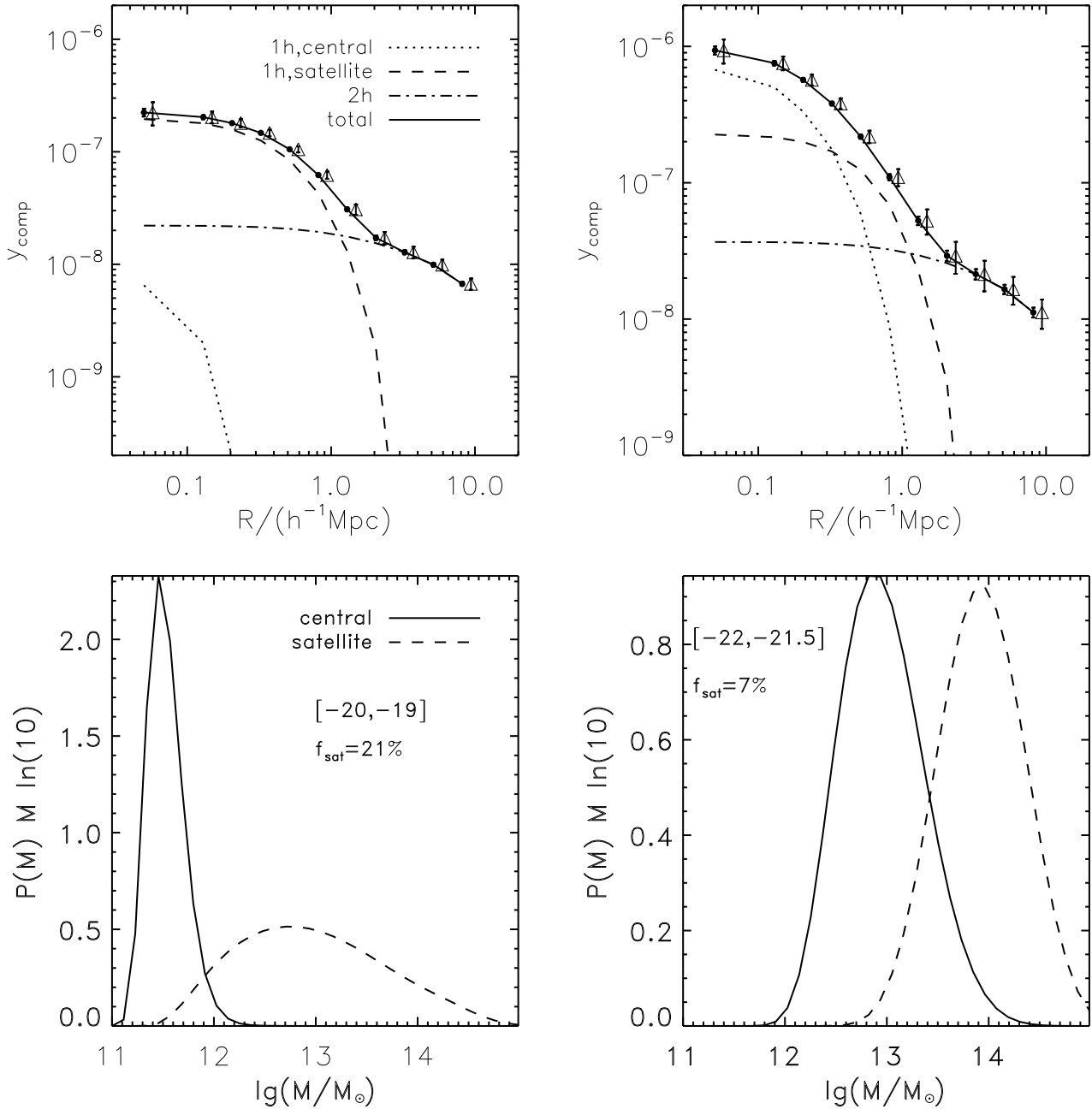}
    \caption{The predicted SZE around galaxies in two different
      luminosity bins (as indicated by the $r$-band absolute
      magnitudes in the low two panels). In the upper panels, the
      predicted $y_{\rm comp}$ parameter is plotted as a function of
      the projected distance to the galaxy. The SZE is decomposed into
      a 1-halo central term (dotted lines), a 1-halo satellite term
      (dashed lines), and a 2-halo term (dash-dotted lines). The total
      signal is shown by the solid line.  The $1\sigma$ expected
      uncertainties from SPT are shown as the errorbars on solid
      circles.  We also plot the $1 \sigma$ confidence level expected
      from the ACT survey as error-bars on triangles.  In the two
      lower panels, the host halo mass distribution of the galaxies
      used to obtain the results shown in the upper two panels are
      shown for central galaxies (solid line) and satellites (dashed
      line).  The distributions are normalized individually.  The
      satellite fraction $f_{\rm sat}$ in each luminosity bin is
      marked in the respective panel.}
    \label{fig:clf}
  \end{center}
\end{figure*}

\begin{figure*}
    \includegraphics[width=\textwidth]{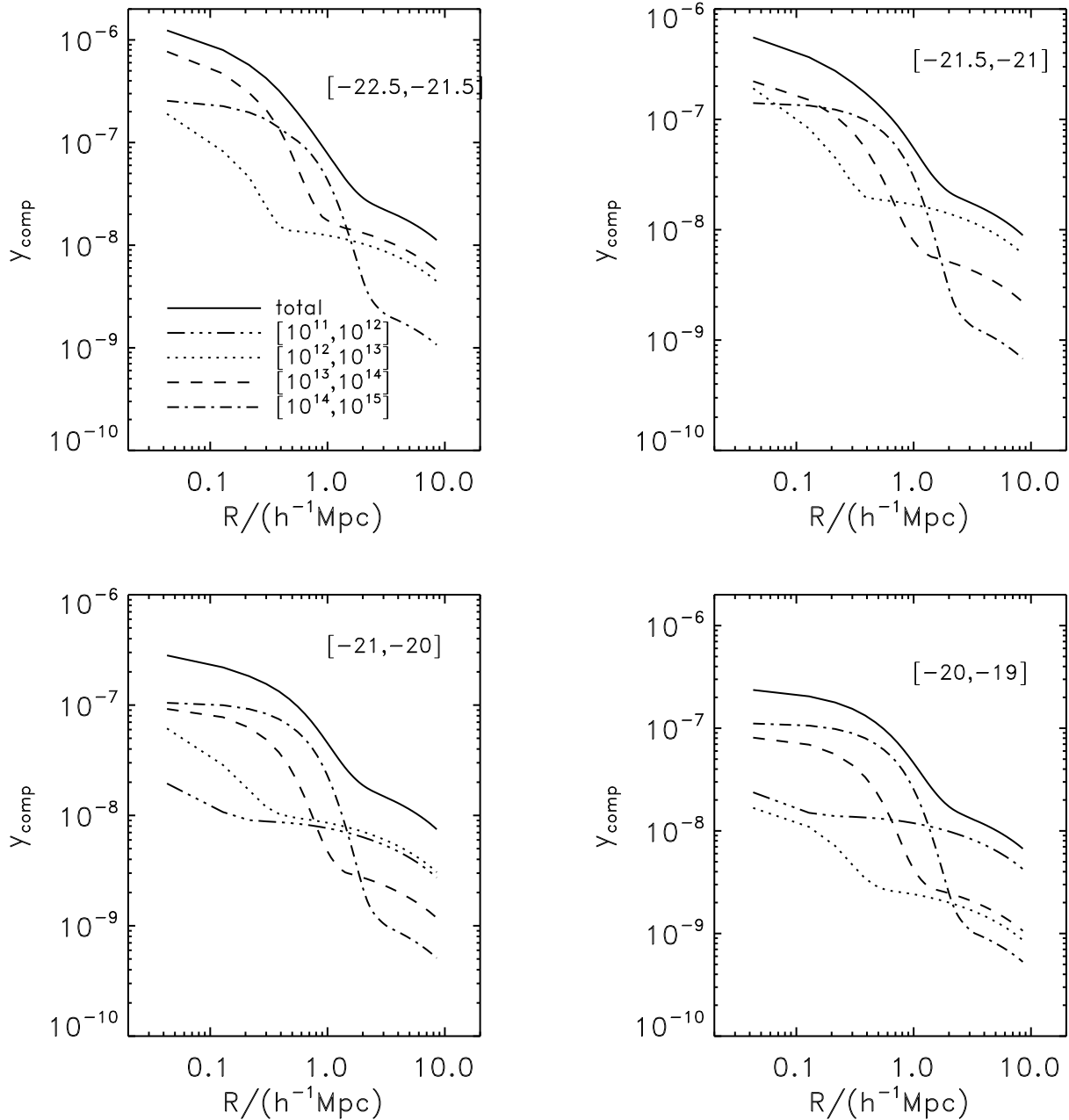}
    \caption{The figure shows how the haloes in different mass
    range contribute to the SZE signal around galaxies that in certain
    luminosity bins. The absolute magnitude range of galaxies are marked on
    top right of each subplot. Different linestyles represent
    contribution from halo of different mass range as shown on top
    left panel.  The unit is in $\msunh$  }
    \label{fig:clf_mass}
\end{figure*}

\begin{figure*}
  \includegraphics[width=\textwidth]{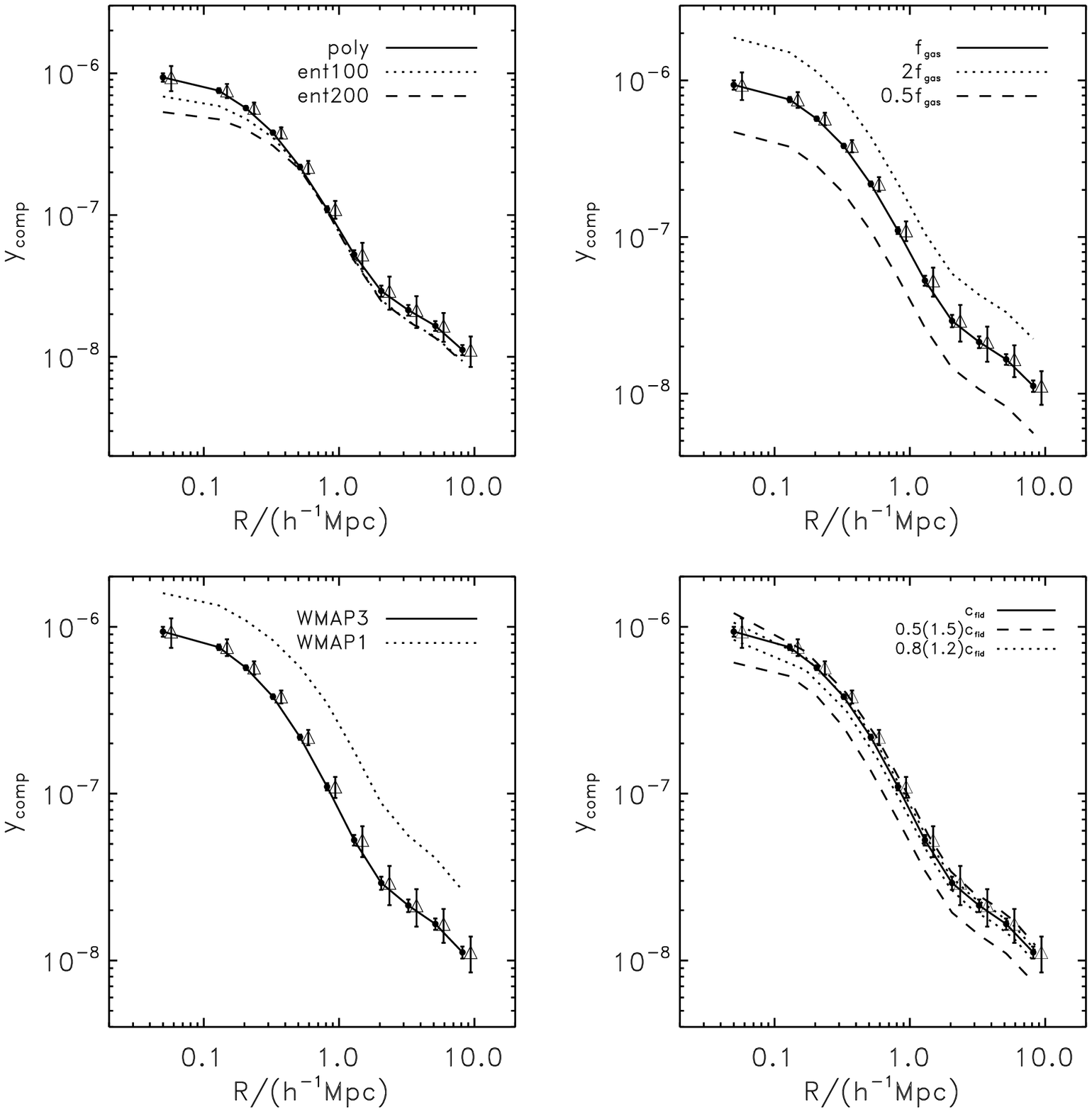}
  \caption{The model dependence of $y_{\rm comp}(R)$ predicted around
    galaxies in the luminosity bin $[-22,-21.5]$.  The $1\sigma$
    expected uncertainties from SPT are shown as the errorbars on
    solid circles. $1 \sigma$ detection limit for ACT survey using
    error bar on triangle.  The upper left panel shows the results for
    different equations of state. The upper right panel shows the
    results for different gas fractions. The lower left panel shows
    the results for two different set of cosmological parameters. And
    the lower right panel shows the effect of changing the halo
    concentration.}
    \label{fig:clf_mod}
\end{figure*}

\subsubsection{The Impact of Mass Uncertainties}

\begin{figure}
\begin{center}
\includegraphics[width=0.5\textwidth]{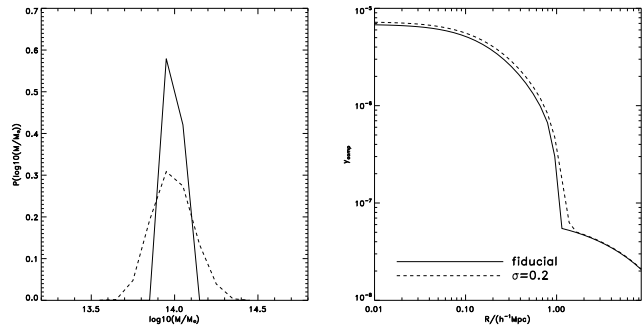}
\caption{The effect of mass assignment errors.  The solid line in the
  left panel shows the original mass distribution in the mass bin
  $[10^{13.9},10^{14.1}]\msunh $ as given by GCY07, while the dashed
  line shows the distribution including the broadening modeled with a
  lognormal distribution with a dispersion of 0.2 dex.  The right
  panel shows the results obtained with these two mass
  distributions. }
\label{fig:scat}
\end{center}
\end{figure}

In the analysis presented above, the calculations are made under the
assumption that there are no errors in the halo mass assigned to each
individual group. However, using mock galaxy redshift surveys, Y07
have shown that the error on the halo mass assigned to an individual
group is of the order of 0.2 to 0.3 dex. To test the importantance of
these errors, we repeat the same exersize as above, but this time
adding a random `error' to the halo mass of each group in the stack,
drawn from a log-normal distribution with a rms of 0.2 dex. The
`perturbed' masses are then used to calculated the SZE signal.
Fig.\ref{fig:scat} shows a comparison between the results thus
obtained with the `perturbed' masses and those obtained assuming the
halo masses are perfectly estimated.  The difference caused by the
mass uncertainties is negligible at large radii (2-halo term). On
small scales, where the 1-halo term dominates, the mass errors cause
an increase in $y_{\rm comp}$ of about $7\%$, and the 1-halo term now
extends to slightly larger radii. This is due to the inclusion of
massive halos in the tail of the mass distribution.  

In real observation, this effect may be quantified by applying the
same analysis to mock group catalogs selected in the same way as the
real catalog.  Using these mock catalogs, one can estimate the
uncertainties in the assigned halo masses, which in turn can be
incorporated in the modeling, similar to what we have done here.

\subsection{The Cross Correlation between Galaxies and the Compton
  Parameter}

Another way to probe the properties of the hot gas in dark matter
halos is to study how the hot gas correlates with galaxies of
different luminosity. One advantage of such analysis is that the
number of galaxies that can be used is large, so that one can stack
the SZE signals around many galaxies, thus achieving high
signal-to-noise. Even more importantly, measurement of the galaxy-SZE
cross correlation does not require a selection of galaxy groups, which
always carries some uncertainties and which typically requires
spectroscopic redshifts for individual galaxies. In fact, as we
demonstrate below, the precision of photometric redshift may be
sufficient to obtain reliable measurements of the galaxy-SZE cross
correlation.  Since photometric redshifts are much easier to obtain
than their spectroscopic counterparts, the galaxy-SZE correlation can
be used to study the hot gas evolution out to higher redshift. The
stacking results can then be interpreted with the CLF discussed in
Section \ref{sec:CLF}
The solid lines in the upper two panels of Fig.~\ref{fig:clf} show
$y_{\rm comp}(R)$ around stacks of galaxies with absolute $r$-band
magnitudes in the ranges $-20 < M_r - 5\log h \leq -19$ (left-hand
panels) and $-21.5 < M_r - 5\log h \leq -22$ (right-hand panels). The
dotted, dashed, and dot-dashed curves show the contributions from the
1-halo central, the 1-halo satellite and the 2-halo terms,
respectively. These are computed using the method described in
Section~\ref{sec:CLF} with the CLF parameters of
\citet{Cacciato09}. Errorbars are obtained using the same method as
described in Section~\ref{sec:spt}, except that rather than using the
numbers of galaxy groups in the GCY07, we use the numbers of galaxies
in the spectroscopic sample of the SDSS DR4 with absolute $r$-band
magnitudes in the respective bins, and with redshifts $0.05 \leq z
\leq 0.20$. The error-bars are estimated using the total independent
angular area of all the angular annuli (corresponding to the bin size
in $R$) around all galaxies in the corresponding luminosity bin, in
comparison to the SPT (ACT) beam size. The lower panels of
Fig.~\ref{fig:clf} show the host halo mass distributions for central
galaxies (solid lines) and satellite galaxies (dashed lines) in the
corresponding magnitude bins.  For clarity, the distributions are
individually normalized, while the satellite fractions in each
luminosity bin are indicated in the corresponding panels.

As one can see, the SZE signal for the brighter sample is dominated by
different contributions on different scales. The 1-halo central term
dominates the signal up to a scale of about $0.3h^{-1}{\rm Mpc}$.  On
scales of $0.3h^{-1}{\rm Mpc} \lta R \lta 2 h^{-1}{\rm Mpc}$ the
signal is dominated by the 1-halo satellite term, while the 2-halo
term dominates on larger scales. For the fainter sample, the situation
is different. Since the host halos of these central galaxies are
mostly low mass halos with $10^{11}\msunh \lta M \lta 10^{12}$
$\msunh$ (see lower left-hand panel), the 1-halo-central term extends
only to relatively small scales and never dominates the signal. The
1-halo satellite term dominates the SZ signal from the center out to
$\sim 1 \mpch$, after which the 2-halo term dominates. This is
important, as it indicates that even very accurate measurements of the
average $y_{\rm comp}(R)$ around relatively faint galaxies yields
virtually no constraints on the properties of hot gas in low mass
haloes (i.e., the haloes that host central galaxies of those
luminosities). Rather, since the SZ signal scales strongly with halo
mass ($Y \propto M^{a}$ with $a \simeq 1.6$), the signal on small
scales is dominated by the 1-halo satellites term, even though
satellites only make up a relatively small fraction ($\sim 21\%$) of
all galaxies in that luminosity bin.  

This is also evident from Fig.~\ref{fig:clf_mass} which shows how
galaxies in halos of different masses contribute to the SZE effect.
Results are shown for four different luminosity bins, as indicated.
In the brightest bin, [-22.5,-21.5], halos in the mass range $10^{13}$
- $10^{14}\, \msunh$ dominate the signal in the inner part, while more
massive halos only dominate the signal between $0.4$ and $1
h^{-1}\Mpc$. This can be understood with the help of
Fig.\ref{fig:clf}. From the inner to outer parts, the signal of the
brightest bin is first dominated by the 1-halo central term and then
by the 1-halo satellite term. Since host halos of satellites on
average are more massive than those of central galaxies of the same
luminosity, the 1-halo satellite term dominates on intermediate
scales. For faint luminosity bins, the SZ signal on small scales
($\lta 1 \mpch$) is always dominated by the most massive halos.
Although only a relatively small fraction of faint galaxies reside in
these massive halos (as satellites) their contribution to the SZ
effect is larger than that from the far more numerous centrals in less
massive halos.

Fig.~\ref{fig:clf_mod} shows the predictions for galaxies in the
[-22.5,-21.5] luminosity bin for different gas equations of state
(upper left-hand panel), different gas mass fractions (upper
right-hand panel), different cosmological parameters (lower left-hand
panel), and different halo concentrations (lower right-hand panel).
Here again we see that the statistical uncertainties expected from a
survey like SPT are much smaller than the differences between
different models, indicating that an analysis along these lines can
put tight constraints on the hot gas properties in dark matter halos
spanning a relatively wide range in masses. 

The lower left-hand panel of Fig.~\ref{fig:clf_mod} shows a comparison
between $y_{\rm comp}(R)$ expected around a stack of galaxies in the
[-22.5,-21.5] luminosity bin in the WMAP1 and WMAP3 cosmologies (see
Table~\ref{tab:cos} for the corresponding cosmological parameters).
In addition to changing the cosmological parameters here we have also
changed the CLF that is used to predict $y_{\rm comp}(R)$. For each
cosmology, the best-fit CLF parameters have been obtained by
\citet{Cacciato09} using the combined constraints from the observed
galaxy luminosity function, the luminosity dependence of galaxy
clustering, and the SDSS group catalogue of \citet{Y07}.  As shown in
\citet{Cacciato09}, both models fit the observed abundance and
clustering of galaxies equally well. However, whereas the WMAP3 model
simultaneously matches the galaxy-galaxy lensing data of
\citet{Mandelbaum06}, the WMAP1 clearly overpredicts the lensing
signal (i.e., the WMAP1 model predicts mass-to-light ratios that are
too high). Hence, galaxy-galaxy lensing can be used to break the
degeneracy between cosmology and halo occupation statistics \citep[see
also][]{Y006,LiRan09}. Interestingly, the results presented here
suggest that the cross correlation between galaxies and the Compton
parameter can also be used to discriminate between these different
models. However, it is clear from Fig.~\ref{fig:clf_mod} that a change
in cosmological parameters has a similar effect as a change in the hot
gas fraction or a change in halo concentration. Hence, it will be
important to have precise constraints on cosmological parameters in
order to use the galaxy-Compton parameter cross correlation to
constrain the properties of hot halo gas. The precision of upcoming
CMB observations, such as Planck, will be able to pin down all
important cosmological parameters to very high accuracy, making the
SZE analysis presented here a powerful tool to study the gas
distribution in dark matter halos.

Finally, for comparison, we also plot the error-bars expected from the
ACT survey. Clearly, the uncertainties in ACT are slightly larger than 
that of SPT and can also tightly constrain the properties of the hot gas
in halos that host these galaxies.

\subsubsection{Photometric Redshifts}

\begin{figure}
    \includegraphics[width=0.5 \textwidth]{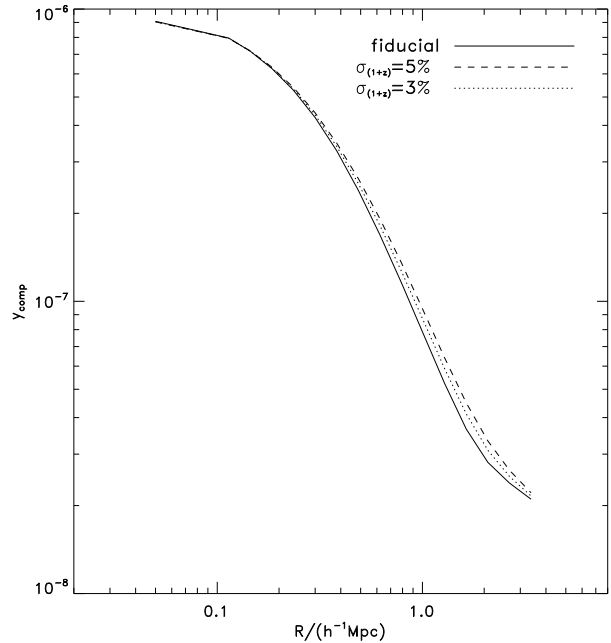}
    \caption{This figure shows how the photo-$z$ error affects the
      observed $y_{\rm comp}$.  The solid line shows the Compton
      parameter $y_{\rm comp}(R)$ in the luminosity bin [-22,-21.5].  The
      dashed and dotted lines show the results of the calculations 
      that include a photo-$z$ error of $5\%$ and $3\%$ in (1+z), 
      respectively.}
    \label{fig:zscat}
\end{figure}

When calculating the cross-correlation between galaxies and hot gas,
we have assumed that the redshifts of the galaxies have no errors.
Although this is a reasonable assumption to make when using
spectroscopic surveys, we now investigate how significantly redshift
errors, such as those present in a photometric redshift survey, impact
on the analysis. Redshift errors introduce errors in both the absolute
magnitudes (luminosities) and the projected distances. In order to
estimate the resulting impact on $y_{\rm comp}(R)$ we add a redshift
error (and its associated luminosity error) to each galaxy in the
$[-21.5, -22]$ luminosity bin of the SDSS DR4 catalog, using a
Gaussian distribution with standard deviations of $(1+z)\times 5\%$
and $(1+z)\times 3\% $.  This roughly mimics the redshift errors
expected from future photometric surveys, such as the LSST. Next we
compute the galaxy-Compton parameter cross correlation using the new
redshifts. Results are shown in Fig.~\ref{fig:zscat}.  The change in
the cross correlation due to photo-z errors of 5\% (3\%) is negligible
at small radii, but increases to $\sim 20\%$ ($\sim 10\%$) at $1 - 2$
$h^{-1} Mpc$.  We also find that, if one can exclude the 20 percentile
of the galaxies with the largest redshift errors, the measurement
errors caused by the photo-$z$ errors shrink to about half this value.
This suggests that the cross correlation analysis considered here can
be extended to photometric surveys, but that precise modeling of the
galaxy-Compton parameter cross correlation may require some treatment
of the photo-$z$ error distribution.

\subsection{Model based on observed X-ray gas profile}

In previous sections, the hot gas profile is modeled based on the 
assumption of hydrostatic equilibrium between the hot gas and 
the dark matter halo. This model is roughly consistent with 
the observed X-ray luminoisty and gas fraction of galaxy 
clusters \citep[e.g.][]{Moodley08}. However, recent work
by \citet{wmap7} showed that the gas profile thus derived 
is different from that derived from X-ray observation
\citep[][A10, hereafter]{arnaud2010}. Since our fiducial 
model is very similar to the KS model (see Section~\ref{sec:physprop}), 
there is the same uncertainty due to the assumed hot gas 
density profile.

\begin{figure}
\includegraphics[width=0.5\textwidth]{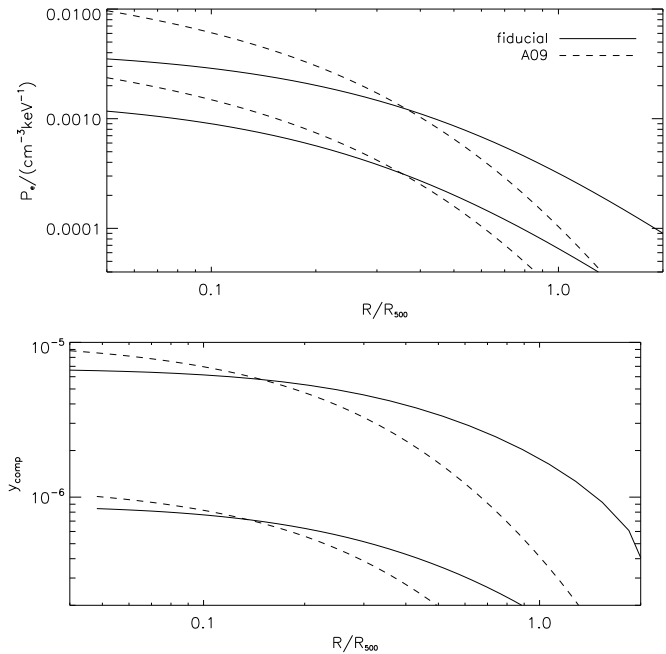}
\caption{The hot gas profile of our fiducial model (solid
    lines) in comparison with that derived in \citep{arnaud2010} 
    (dashed lines). The upper panel shows the hot gas pressure 
    profile as a function of radius. The bottom panel shows the 
    Compton $y$-parameter profile. In each panel, the upper 
    set of lines are for the halo of  $10^{14}\msunh$ and the 
    lower set of lines for halo of $10^{13}\msunh$. The horizontal 
    axis shows radii scaled by the corresponding $R_{500}$.}
\label{fig:a09_P}
\end{figure}

In Fig.\ref{fig:a09_P}, we compare our hot gas pressure 
profile with that derived from A10, with the latter obtained 
using the formula given in the Appendix of \citet{wmap7}. 
The horizontal axis is scaled by $R_{500}$, the radii within
which the mean density is 500 times the critical density. 
As in \citet{wmap7}, we also find that the pressure profile 
predicted by our model is flatter and more extended.
This difference causes a significant difference in the 
predicted SZE at large radii. 

One possibility to alleviate the tension between the two 
models is to change the gas fraction and boundary condition 
in our model. In our model, the boundary condition 
is set by $T(r_{\rm vir})=T_{\rm vir}$, but X-ray observations
indicate that the temperature drops more rapidly in the outer part 
of the halo. For some halo massed, the slope of the X-ray 
derived pressure profile can be fitted by decreasing the 
temperature at virial radius. However, in hydrostatic equilibrium 
the low boundary temperature results in a higher inner 
gas density, and the amplitude of SZE at small scales 
becomes higher than that based on X-ray observations. 
For a halo of $10^{14} \msunh$, we can decrease the discrepancy 
between our fiducial profile and X-ray profile by 
setting $T(r_{\rm vir})=0.3T_{\rm vir}$ and the gas to total
fraction $f_{\rm gas}=0.08$. However, we cannot find a single
boundary condition that can fit the X-ray profile for all 
halo masses. Because of these uncertainties, the boundary 
condition should be treated as a free parameter when studying
the hot gas properties of groups with future SZE observation.

Observationally, X-ray observations lack the sensitivity 
to reliably probe hot gas beyond $R_{500}$, 
because the X-ray luminosity scales with the square of 
the gas density. In A09, the pressure profile is derived 
using REXCESS data \citep{Bohringer2007}, where X-ray 
measurements are limited within $R_{500}$. Thus, the pressure 
profile at larger radii is still uncertain. 

\begin{figure}
\includegraphics[width=0.5\textwidth]{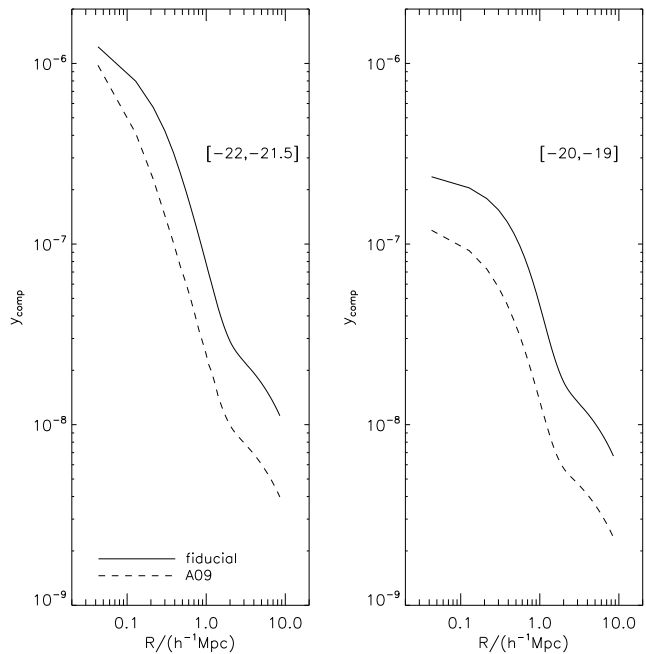}
\caption{The $y_{\rm comp}(R)$ around galaxies in
  luminosity bins $[-22,-21.5]$ (left) and $[-20,-19]$ (right) 
  predicted by our fiducial model (solid lines) 
  in comparison with that predicted with the formula derived 
  from \citep{arnaud2010} (dashed lines).}
\label{fig:a09_clf}
\end{figure}

In order to test how the uncertainties in the hot gas pressure 
profile affect our prediction of the SZE, we compare 
in Fig.\ref{fig:a09_clf} the cross-correlation between 
the SZE and galaxies predicted by our fiducial model 
with that predicted by the X-ray observation-based model. 
The predictions of our fiducial model is a factor
of 2 to 3 larger. This indicates that the observation 
of the cross-correlation between SZE and galaxies can be 
used to distinguish these two models.

\section{Conclusions}
\label{sec:sum}

In this paper we have used simplified models to demonstrate the
potential of using the SZE to probe the hot gas expected to be
associated with dark matter halos. We have shown that by stacking SZE
maps expected from the Planck, ACT and/or SPT surveys, one can probe
the hot gas properties in galaxy groups with halo masses down to $\sim
10^{13}\msunh$. The SZE for halos with similar masses are examined in
terms of halo models. Splitting the total SZE signal into 1-halo and
2-halo terms, we have shown that in high-mass halos the 1-halo term
dominates the signal within the virial radius; in low-mass halos the
2-halo term is also significant on small scales. The model predictions
are sensitive to the amount of hot gas assumed to be in halos, and our
results show that a SPT-like survey can provide a stringent constraint
on the hot gas fraction for halos with masses $M\ga
10^{13}\msunh$. Furthermore, such observations can also provide
stringent constraints on the equation of state of the hot gas as well
as the concentrations of dark matter halos.

Using WMAP-7-year data, \citet{wmap7} stacked clusters in X-ray
catalog and compared the resulting SZE profile with the prediction of
the KS model\citep{Kom01} that uses a polytropic gas equation
of state very similar to the polytropic model considered here.
They found that this model over-predicts
the SZE by a factor of 1.5 to 2. In a recent study using SPT data,
\citet{lueker09} also found that the observed SZE power spectrum is
lower than that expected from polytropic model.  However, as we have
shown above, entropy injection can reduce the amplitude of the SZE by
a factor of 1.5 at the inner parts of halos with masses
$10^{14}\msunh$, suggesting that non-gravitational heating may have
played an important role.

We have also explored the idea of using the cross correlation between
hot gas and galaxies of different luminosity to probe the hot gas in
dark matter halos with the use of the CLF. Since the number of
galaxies that can be used is large, especially in the forthcoming
large photometric surveys where accurate photometric redshifts of
individual galaxies can be obtained, the measurement can be made to
high precision. Our results show that, with the help of CLF modeling,
one can constrain the hot gas profile in halos with masses down to
$10^{13}\msunh$. Cosmological parameters, especially the value of
$\sigma_8$, can affect galaxy - hot gas cross-correlation. Thus, in
order to use the observed SZE to constrain the hot gas properties in
dark matter halos, precise constraints on cosmological parameters are
required.

In summary, the upcoming SZE surveys are expected to provide an
important avenue to study the hot gas properties in dark matter
halos. Combining with large optical galaxy surveys and using the
stacking method proposed here, the SZE to be observed in forthcoming
surveys will allow us to study in great detail the hot gas properties
in galaxy groups where most galaxies reside. This, in turn, will shed
new light on how galaxy form and evolve in dark matter halos.

\section*{Acknowledgments}

RL is supported by the National Scholarship from China Scholarship
Council. Part of the computation was carried out on the SGI Altix 330
system at the Department of Astronomy, Peking University. HJM
acknowledges the support of NSF 0908334 and NSF 0607535.  ZHF is
supported by NSFC of China under grants 10373001,10533010, 10773001
and 973 program No.2007CB815401. We thank Heling Yan for useful
discussion.  

\bibliography{sz_paper} 
\bibliographystyle{mn2e}

\appendix 

\section{Angular power spectrum and SZE map}
We write the SZE power spectrum, $P_{\rm sz}(k,z)$, as 
\begin{equation}
P_{\rm sz}(k ,z)=P_{\rm sz}^{\rm 1h}(k , z)+P_{\rm sz}^{\rm 2h}(k , z) \,,  
\end{equation}
where the 1-halo contribution is 
\begin{equation}
P_{\rm sz}^{\rm 1h}(k,z) = I^{2}_{1,\rm sz}(k,z), 
\end{equation}
and the 2-halo contribution is
\begin{equation}
P_{\rm sz}^{\rm 2h}(k,z) = P_{\rm lin}(k,z)  I_{2,\rm sz}^2\,.
\end{equation}
Here
\begin{equation}
I_{1,\rm sz}^{2}(k,z) = \int_{M_{\rm min}}^{M_{\rm max}} n(M,z) \,
{\tilde u}_{p}^{2}(k|M,z) \, {\rm d}M \,. 
\end{equation}
and
\begin{equation}
I_{2,\rm sz}(k,z) = \int_{M_{\rm min}}^{M_{\rm max}} n(M,z) \,
{\tilde u}_{p}(k|M,z) \, b(M,z)\, {\rm d}M\,. 
\end{equation}
Observationally, one typically measures the angular power spectrum,
which is the integration of the power spectrum along the
line-of-sight. To derive the angular power spectrum, we expand the CMB
temperature fluctuations due to the SZE in spherical harmonics:
\begin{equation}
\frac{\Delta T}{T_{{\rm CMB}}} (\hat{\bf n}) = \sum_{lm} 
\left(\frac{\Delta T}{T_{{\rm CMB}}} \right)_{lm} Y_{lm}(\hat{\bf  n})
\end{equation}
The angular power spectrum can then be written as:
\begin{equation}
\left\langle \left( \frac{\Delta T}{T_{\rm CMB}}\right)_{lm}
\left(\frac{\Delta T}{T_{\rm CMB}}\right)_{l'm'}\right\rangle
=C_{l}^{\rm sz}\delta_{l l'}\delta_{m m'} \,.
\end{equation}
Using the Limber approximation \citep[][]{Limber53}, we have
\begin{equation}\label{eq:cl}
C_{l}^{\rm sz}=\int^{z_{\rm max}}_{z_{\rm min}} dz 
\frac{{\rm d}^{2}V}{{\rm d}z {\rm d}\Omega}  P_{l, \rm sz}(l,z)\,,
\end{equation}
where ${\rm d}^{2}V/{\rm d}z {\rm d}\Omega$ is the comoving volume per
unit redshift per unit solid angle, and the projected power spectrum,
$P_{l,\rm sz}(l,z)$, is related to the 3-dimensional power spectrum
via
\begin{equation}
P_{l,\rm sz}(l,z)=\left[\frac{W_{\rm sz}(z)}
{D_{\rm c}(z)}\right]^{2}\,\left[I_{1,\rm sz}^{2}(l,z)
+I_{2,\rm sz}^2(l,z)\right]\,,
\end{equation}
with
\begin{equation}
W_{\rm sz}(z) = \Theta_{\nu}\frac{\sigma_{T}}{m_{\rm e}} 
\frac{k_{\rm B}}{a^2(z)}\,.
\end{equation}
Here $D_{\rm c}(z)$ is the comoving distance from 
redshift $z$, and $a(z)$ is the scale factor.

\begin{figure}
\begin{center}
\includegraphics[width=0.5\textwidth]{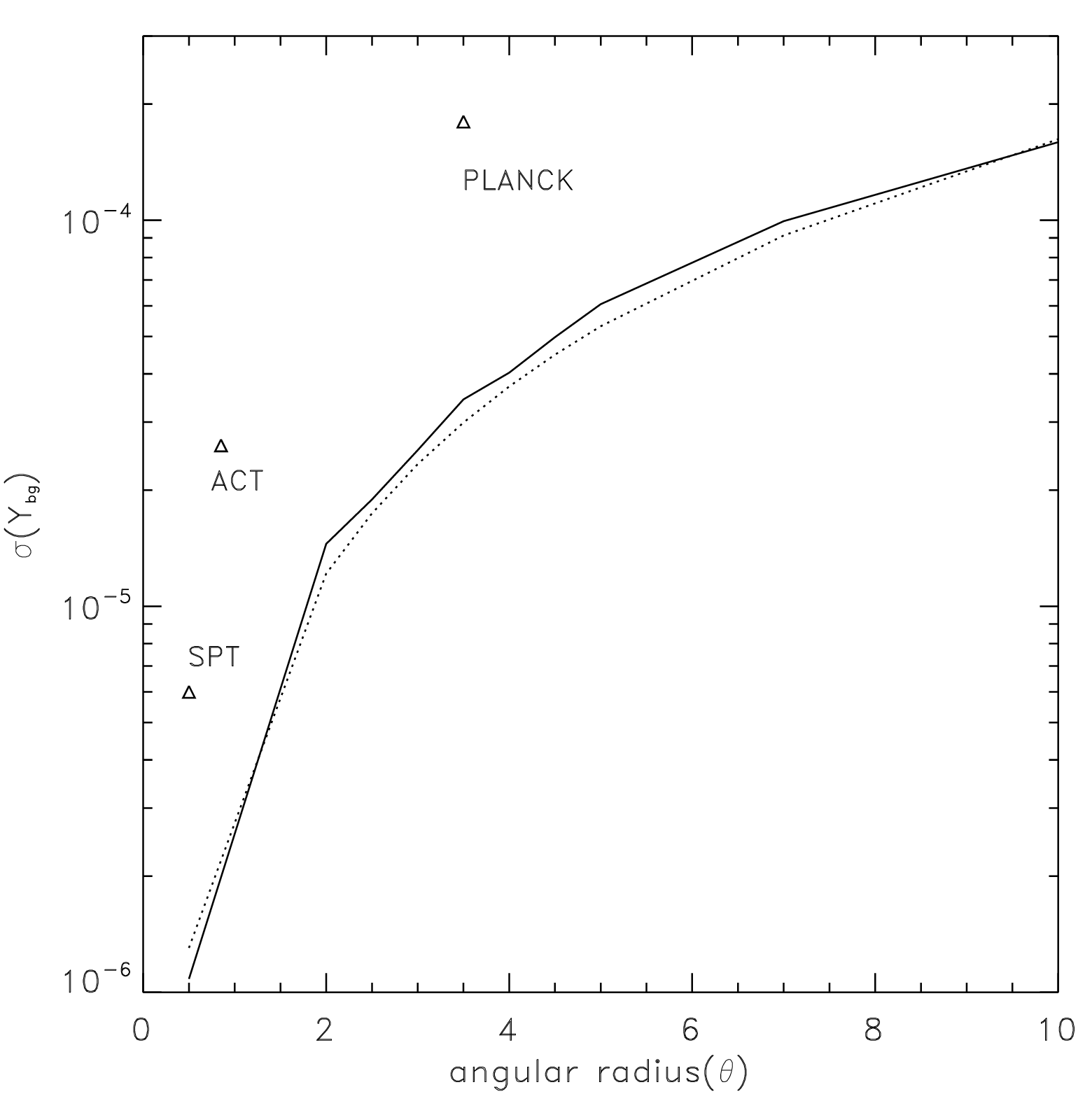}
\caption{The fluctuation of the SZE background as a function of
  angular radius $\theta$. The solid line shows our result, and the
  dotted line shows the fitting formula of \citet{shaw08}.  The
  triangles show the instrumental noise levels of SPT (150GHz), ACT
  triangles show the instrumental noise levels of SPT (143GHz), ACT
  (148 GHz) and Planck (150GHz), respectively.}
\end{center}
\label{fig:bg}
\end{figure}

We carry out the integration using Eq.(\ref{eq:cl}) from redshift
$z_{\rm min} = 0$ to $z_{\rm max} = 5$. For the mass limits, we use
$M_{\rm min} = 5 \times 10^{11} \msunh$ and $M_{\rm max} = 5 \times
10^{15} \msunh$.  We then use the angular power-spectrum as an input
to the synfast subroutine in the HEALPIX package \citep[][]{Gorski05}
in order to generate an all sky SZE map. In Fig.\ref{fig:bg}, we show
the standard deviation of the background SZE flux measured within an
angular radius $\theta$ (solid line) and compare it with the
instrumental noise levels of SPT, ACT and Planck. The background noise
is much smaller than the instrumental noise. We also compare our
result with the fitting formula given by \citet{shaw08} (dotted line)
who use the background SZE noise measured from a light cone simulation
with cosmological parameters similar to ours. Note the good agreement
between the two.

\end{document}